\documentclass[lettersize,journal]{IEEEtran}
\usepackage{amsmath,amsfonts}
\usepackage{algorithm}
\usepackage{algorithmicx}
\usepackage[]{algpseudocode}
\usepackage{soul}
\usepackage{bm}
\usepackage{caption}

\usepackage{array}
\usepackage[caption=false,font=normalsize,labelfont=sf,textfont=sf]{subfig}
\usepackage{textcomp}
\usepackage{multirow}
\usepackage[table]{xcolor}
\usepackage[colorlinks=true,linkcolor=black,citecolor=black,urlcolor=black,anchorcolor=black,runcolor=black,pdfborder={0 0 0}]{hyperref}

\makeatletter
\newcommand{\labeltext}[1]{%
  #1%
  \protected@xdef\@currentlabel{\unexpanded{#1}}%
}
\makeatother

\definecolor{myblue}{RGB}{25, 50, 80} 
\makeatletter
\AtBeginDocument{%
  \hypersetup{citecolor=myblue, linkcolor=myblue}%
}
\makeatother

\usepackage{stfloats}
\usepackage{url}
\usepackage{verbatim}
\usepackage{amssymb}
\usepackage{pifont}
\usepackage{graphicx}
\usepackage{booktabs}
\usepackage{cite}
\usepackage{xcolor}
\begin{document}

\title{HGraphScale: Hierarchical Graph Learning for Autoscaling Microservice Applications in Container-based Cloud Computing}

\author{Zhengxin Fang,~\IEEEmembership{Graduate Student Member,~IEEE,} Hui Ma,~\IEEEmembership{Senior Member,~IEEE,} Gang Chen,~\IEEEmembership{Senior Member,~IEEE,} and Rajkumar Buyya,~\IEEEmembership{Fellow,~IEEE}
\thanks{Z. Fang, H. Ma and G. Chen are with the School of Engineering and Computer Science \& Centre for Data Science and Artificial Intelligence, Victoria University of Wellington, Wellington, New Zealand. E-mail: \{zhengxin.fang, hui.ma, aaron.chen\}@ecs.vuw.ac.nz. }
\thanks{R. Buyya is with the School of Computing and Information Systems, the University of Melbourne, Melbourne, Australia. Email: rbuyya@unimelb.edu.au}}

\markboth{Journal of \LaTeX\ Class Files,~Vol.~14, No.~8, August~2021}%
{Shell \MakeLowercase{\textit{et al.}}: A Sample Article Using IEEEtran.cls for IEEE Journals}


\maketitle

\begin{abstract}
Microservice architecture has become a dominant paradigm in application development due to its advantages of being lightweight, flexible, and resilient. Deploying microservice applications in the container-based cloud enables fine-grained elastic resource allocation. Autoscaling is an effective approach to dynamically adjust the resource provisioned to containers. However, the intricate microservice dependencies and the deployment scheme of the container-based cloud bring extra challenges of resource scaling. This article proposes a novel autoscaling approach named HGraphScale. In particular, HGraphScale captures microservice dependencies and the deployment scheme by a newly designed hierarchical graph neural network, and makes effective scaling actions for rapidly changing user requests workloads. Extensive experiments based on real-world traces of user requests are conducted to evaluate the effectiveness of HGraphScale. The experiment results show that the HGraphScale outperforms existing state-of-the-art autoscaling approaches by reducing at most 80.16\% of the average response time under a certain VM rental budget of application providers.

\end{abstract}

\begin{IEEEkeywords}
Autoscaling, Microservice application, Container-based cloud, Graph neural network, Deep reinforcement learning
\end{IEEEkeywords}

\section{Introduction}\label{section:introduction}
\labeltext{Microservice applications is marking a paradigm shift in how software systems are designed and managed~\cite{blinowski2022monolithic}. These modern applications are composed of lightweight and scalable microservices, which improve scalability, agility, and resilience~\cite{berry2024worth}.}\label{change:R2C3-8} \labeltext{Each microservice is instantiated by one or more containers.}\label{change:R2C6-1} Building on this paradigm, cloud computing serves as a critical enabler for hosting and scaling microservice applications~\cite{al2017elasticity}.

\labeltext{The dynamic resource adjustment in cloud computing, known as \emph{autoscaling}~\cite{bai2024drpc,wen2025statuscale,jeong2025autoscaling,li2025astra}, empowers microservice applications to efficiently handle fluctuating
user requests~\cite{wang2022performance} by leveraging \emph{horizontal scaling} and \emph{vertical scaling} techniques.}\label{change:R2C10-1} Horizontal scaling creates or deletes raplicas of containers, while vertical scaling adjusts the resources (e.g., CPU) provisioned to individual container. 

\begin{figure}[htb!]
    \centering
    \includegraphics[width=0.8\linewidth]{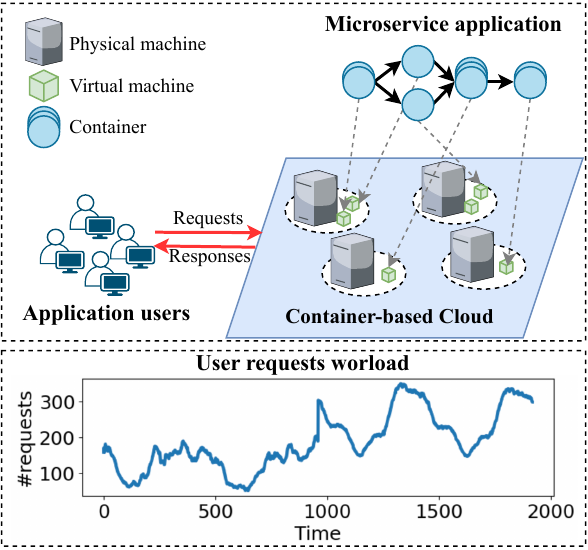}
    \caption{Microservice application deployed in the container-based cloud with fluctuating user requests.}
    \label{fig:AMC}
\end{figure}

The effectiveness of autoscaling is further enhanced by the container-based cloud~\cite{tan2020cooperative,fang2023energy,hamzaoui2024topical,fang2024leveraging}, which offers fine-grained resource allocation tailored to dynamic workloads. As illustrated in Fig.~\ref{fig:AMC}, microservice applications deployed in the container-based cloud follow a hierarchical structure: containers are hosted within Virtual Machine instances (VMs), which, in turn, are deployed on Physical Machine instances (PMs). 

Within the container-based cloud, the Quality of Service (QoS) of microservice applications, such as their average response time~\cite{cheng2023proscale}, depends on the number of containers and the resources provisioned to them~\cite{xie2024pbscaler,bai2024drpc,shi2023auto}. To maintain high QoS under fluctuating user requests workload (as shown in Fig.~\ref{fig:AMC}), this article investigates the critical problem of \emph{\underline{A}utoscaling \underline{M}icroservice applications in the \underline{C}ontainer-based cloud}, referred to as the AMC problem in the remaining of this paper.

In the AMC problem, provisioning excessive resources to containers may help meet the service level objectives (SLOs) of application providers but often leads to significant resource wastage~\cite{baarzi2021showar,park2021graf}. \labeltext{The cloud resources wastage increases cost for application providers due to unnecessary VM rentals}\label{change:R2C3-1}~\cite{huang2022cost,shi2023auto,xie2024pbscaler}. Moreover, the intricate dependencies between containers introduce significant challenges in autoscaling~\cite{meng2022hra,gari2024online}. Additionally, the finite resource capacities of PMs in the container-based cloud exacerbate the complexity of provisioning the right amount of resources to containers. 

Given the above challenges, an effective autoscaling approach is essential to enhance the QoS of applications while adhering to a defined cost budget. However, many existing approaches rely on simple threshold-based mechanisms, such as Amazon auto-scaling service~\cite{AWSAutoScaling2022} and Horizontal Pod Auto-scaler (HPA)~\cite{Burns2019Kubernetes}. These methods make scaling actions based on a pre-defined threshold. Nonetheless, manually selecting a threshold for changing workload is non-trivial. An inappropriate threshold can easily result in under-provisioning, leading to QoS degradation, or over-provisioning, causing unnecessarily high costs~\cite{imdoukh2020machine}.

To address the limitations of threshold-based approaches, \labeltext{Deep Reinforcement Learning (DRL) is a promising approach for autoscaling~\cite{bai2024drpc,meng2023deepscaler,meng2022hra,liang2026autoscaling}. It can automatically learn generalizable scaling policies that adapt to dynamic environments.}\label{change:R2C3-2} DRL-based methods employ deep neural networks, such as Graph Neural Networks (GNNs), to obtain container embeddings. \labeltext{These embeddings capture implicit characteristics and complex dependencies among containers, which are then used to guide scaling decisions.}\label{change:R2C3-3} However, it is not intuitive to design an effective embedding learning approach for the AMC problem since the container-based cloud is a rather complex system. Two major issues have not been addressed by existing studies.

\labeltext{First, existing DRL-based autoscaling approaches~\cite{shi2023auto,bai2024drpc,meng2022hra} do not explicitly consider the \emph{deployment scheme} of the container-based cloud~\cite{cheng2024geoscale}. The deployment scheme specifies the hierarchical mapping of system components: containers are assigned to virtual machines (VMs), and virtual machines are in turn assigned to physical machines (PMs).}\label{change:R2C6-2} These approaches only focus on the features of individual components, such as containers, VMs and PMs. The deployment scheme is critical because different deployment schemes result in varying resource constraints (e.g., the capacities of VMs and PMs), which directly influence the effectiveness of scaling actions. Ignoring this factor can lead to suboptimal scaling actions and resource utilization.

Second, GNNs employed in existing autoscaling approaches~\cite{meng2023deepscaler,tong2023gma} aggregate information in a \emph{flat} way. \labeltext{In this flat structure, containers, VMs and PMs are modeled in a single layer. As a result, long-range dependencies (e.g., between containers on different PMs) require many message-passing steps to capture.}\label{change:R3C15} \labeltext{Studies~\cite{ma2022hierarchical,wu2020learning,zhong2023hierarchical} have shown that the flat GNN structure cannot effectively capture long-range dependencies for learning node embeddings.}\label{change:R2C3-7} This issue becomes more aggravated in the AMC problem with the increasing number of containers, VMs and PMs in the container-based cloud.

To address the above issues, this paper focuses on designing a novel embedding learning approach to improve the performance of the DRL-based autoscaling approach. \labeltext{For this purpose, we construct a three-layer hierarchical graph. It models both the dependencies among containers and the deployment scheme.}\label{change:R2C3-4} The hierarchical structure of this graph, from bottom to top, is: \emph{PM layer}, \emph{VM layer} and \emph{container layer}. Then, we design a novel Hierarchical Graph Neural Network (HGNN) to solve the issue of long-range information aggregation.

HGNN is a solution for the issue of long-range information aggregation~\cite{ma2022hierarchical,zhong2023hierarchical}. \labeltext{However, existing HGNN approaches~\cite{ma2022hierarchical,ying2018hierarchical} mainly learn whole-graph embeddings by aggregating information in a fine-to-coarse manner.}\label{change:R2C3-5} These approaches are effective for tasks that require holistic graph representations. However, it is not well-suited for the AMC problem, where precise scaling actions depend on embeddings at the granularity of individual containers rather than the entire graph.

To fill this gap, we proposed a \labeltext{\emph{Cloud-oriented Hierarchical Graph Neural Network (CHGNN)}, which is an HGNN designed to effectively learn container embeddings from cloud environment.}\label{change:R2C6-3} \labeltext{Unlike traditional methods, CHGNN first aggregates information locally within lower-layer nodes and then propagates it to higher layers.}\label{change:R2C3-6} This \emph{bottom-up information aggregation mechanism} establishes shortcut connections~\cite{ma2022hierarchical,wu2020learning,he2016deep} between distant nodes in the graph, enabling effective processing of global context. Consequently, CHGNN can capture comprehensive global information from the container-based cloud for the container layer. This mechanism not only represents a departure from existing HGNN paradigms but also delivers a more precise and scalable solution for the AMC problem.

Through developing CHGNN, this paper makes the following main contributions:

\begin{itemize}
    \item  We represent the container-based cloud as a three-layer hierarchical graph. Meanwhile, we design a novel HGNN, i.e., CHGNN, to learn container embedding from the hierarchical graph. To our knowledge, this is the first work to learn embedding for autoscaling using HGNN, allowing to make more effective scaling actions for the AMC problem than existing approaches.

    \item We propose a novel bottom-up information aggregation mechanism for CHGNN to effectively capture thorough global information from the container-based cloud. This mechanism provides an accurate and scalable autoscaling solution to the AMC problem.

    \item We propose a novel DRL-based autoscaling approach that leverages CHGNN with a bottom-up information aggregation mechanism to effectively learn container embeddings. In addition, a newly designed \emph{scaling policy network} is employed to make scaling decisions. We name this autoscaling approach as HGraphScale. Experiment results based on real-world traces indicate that HGraphScale can outperform five state-of-the-art autoscaling approaches.
\end{itemize}

The rest of this article is organized as follows. Section~\ref{section:related} presents the literature review of existing autoscaling approaches. Section~\ref{section:problem} presents formal problem definitions of the problem. Section~\ref{section:HGraphScale} gives details of HGraphScale for autoscaling. The experiment designs, results and further analysis are shown in Section~\ref{section:experiment}. At last, Section~\ref{section:conclusion} makes conclusions and gives potential future directions.

\section{Related Work}\label{section:related}
\labeltext{In this section, we review existing autoscaling approaches for microservice applications.}\label{change:R2C8-1}

\subsection{Heuristic-based Autoscaling}
AWS-Scale~\cite{AWSAutoScaling2022} and Horizontal Pod Auto-scaler (HPA)~\cite{Burns2019Kubernetes} are autoscaling techniques that rely on manually determined thresholds. For example, the resources provisioned to containers are increased if the resource utilization is higher than a given threshold; otherwise, it decreases the resources provisioned to containers. However, manually designing a threshold for changing workload is challenging.

To address the above issue, some heuristic-based autoscaling approaches are proposed to make scaling actions based on predicted future workload. 
ProScale~\cite{cheng2023proscale} is a proactive autoscaling method that leverages the accurate and fast Simple Moving Average (SMA) to predict future request workloads. Then, the resource adjustment of containers is based on a greedy method.  PBScaler~\cite{xie2024pbscaler} is proposed to detect the bottleneck microservices in an application. Subsequently, a genetic algorithm is applied to decide the number of containers required by bottleneck microservices. \labeltext{StatusScale~\cite{wen2025statuscale} is a status-aware autoscaling approach that selects appropriate autoscaling strategies for 
resource scheduling based on load status.}\label{change:R2C10-2}

The above autoscaling methods require substantial human efforts to design the heuristics or fine-tune the thresholds. Meanwhile, the heuristics methods exhibit poor generalization ability in dynamically changing environments~\cite{yang2022dual}.

\subsection{Reinforcement Learning-based Autoscaling}
Existing studies~\cite{shi2023auto,bai2024drpc,meng2022hra,gari2022q,gari2024online,zhang2020sarsa} have shown that the RL-based autoscaling methods can effectively adjust the resource allocation to handle the changing workload. For example, A-SARSA is proposed~\cite{zhang2020sarsa} to combine neural network based workload prediction and the SARSA algorithm~\cite{watkins1992q} to make scaling actions based on predicted workload. 

A Q-learning based autoscaling approach~\cite{gari2022q} is proposed for workflow autoscaling, which considers the workflow structure when making scaling actions. \cite{gari2024online} further compared the performance of Q-learning and SARSA for workflow autoscaling, considering the workflow structures. Their results show that SARSA can achieve significantly better performance in many scenarios compared to Q-learning.

The above approaches use table-based RL techniques, struggling to handle high-dimensional state spaces. To tackle this limitation, the DRL-based autoscaling approach has been gaining more attention in recent years. For instance, a Deep Q-Newtork (DQN)~\cite{mnih2015human} based autoscaling method, called HRA~\cite{meng2022hra}, is proposed to make holistic autoscaling actions for microservice applications. Similarly, DeepScale~\cite{shi2023auto} integrates DQN and heuristics methods to make scaling actions for applications. DRPC~\cite{bai2024drpc} is a TD3~\cite{fujimoto2018addressing} based DRL approach to make scaling actions based on embedding learned by multiple distributed neural networks. ASTRA~\cite{li2025astra}, a recently introduced approach, leverages an adversarial DRL algorithm for autoscaling.

The above DRL-based approaches fail to explicitly consider the deployment scheme, which impacts scaling actions. This hinders the effectiveness of these methods in addressing the AMC problem. 

\subsection{Graph Neural Network-based Autoscaling}\label{subsection:related-gnn}
\labeltext{Beyond heuristic and DRL-based autoscaling methods, GNN-based approaches have emerged as a popular solution for autoscaling. For instance, DeepScaler~\cite{meng2023deepscaler} is proposed to estimate resource utilization by GNN, which is further utilized to guide the autoscaling decisions. GRAF~\cite{park2021graf,park2024graph} is proposed to predict tail latency of microservice applications. The predicted latency is leveraged for proactive autoscaling decision making. AGQ~\cite{liang2026autoscaling} is an autoscaling approach that utilizes a GNN-based resource usage predictor, which directly informs the autoscaler’s scaling decisions.}\label{change:R3C16-1}

\labeltext{In a summary, existing GNN-based autoscaling approaches are designed for prediction tasks, either forecasting resource usage or predicting latency. Such prediction tasks require large datasets for training, and these GNNs do not account for the deployment scheme in the cloud environment.}\label{change:R3C16-2}

\subsection{Summary}
To address the above limitations of existing autoscaling approaches, this article proposes HGraphScale, a novel DRL-based autoscaling approach that incorporates a newly designed GNN and information aggregation mechanism. The details of comparison between HGraphScale and other DRL-based and GNN-based autoscaling approaches are shown in TABLE~\ref{table:comparison}

\begin{table}[htb!]
    \centering
    \caption{Comparison of HGraphScale with DRL-based and GNN-based autoscaling approaches}
    \scalebox{0.79}{\begin{tabular}{|l|c|c|c|c|c|c|c|}
        \hline
        \textbf{} & \multicolumn{7}{c|}{\textbf{Approaches}} \\
        \cline{2-8}
        & \cite{liang2026autoscaling} & \cite{gari2022q} & \cite{meng2022hra} & \cite{shi2023auto} & \cite{bai2024drpc} & \cite{gari2024online} & HGraphScale  \\
        \hline
        \textbf{Vertical scaling} & &  &  & \checkmark & \checkmark &  & \checkmark  \\
        \hline
        \textbf{Horizontal scaling} & \checkmark & \checkmark & \checkmark & \checkmark & \checkmark & \checkmark & \checkmark  \\
        \hline
        \textbf{QoS improvement} & \checkmark & \checkmark & \checkmark &  & \checkmark & \checkmark & \checkmark  \\
        \hline
        \textbf{Cost saving} & \checkmark &  & \checkmark & \checkmark &  & \checkmark & \checkmark  \\
        \hline
        \textbf{High dimensional state} &  &  & \checkmark & \checkmark & \checkmark &  & \checkmark  \\
        \hline
        \textbf{Microservice dependency}  & \checkmark &  &  &  &  & \checkmark & \checkmark  \\
        \hline
        \textbf{Deployment scheme} &  &  &  &  &  &  & \checkmark  \\
        \hline
    \end{tabular}}
    
    \label{table:comparison}
\end{table}

\section{Problem Description}\label{section:problem}
\labeltext{In this section, we formulate the problem of Autoscaling Microservice application in the Container-based cloud (the AMC problem)}\label{change:R2C8-2} The AMC problem aims to elastically provision or deprovision resources to containers in response to dynamically changing workload, optimizing the QoS under a cost budget. TABLE~\ref{table:notations} summarizes the notations used in this article.

\begin{table}[htb!]
\caption{Notations used throughout this article}
\label{table:notations}
\begin{center}
    \begin{tabular}{|p{1.4cm}|p{6.55cm}|}
    \hline
    \textbf{Notation} & \textbf{Meaning}            \\ 
    \hline
    $ms_i$ & a microservice type $i$ \\
    \hline
    $Con_i^j$ & a container that $ms_i$ is instantiated \\
    \hline
    $t_i$ & a task in a user request that is executed by $ms_i$ \\
    \hline
    $e^{app}_{i,j}$ & execution dependency between a pair of adjacent microservices \\
    \hline
    $e^{wf}_{i,j}$ & execution order of two tasks in a workflow \\
    \hline
    $t_{start}$ & dummy start task of a request \\
    \hline
    $t_{end}$ & dummy end task of a request \\
    \hline
    $et_i$ & execution time of $t_i$ with one vCPU \\
    \hline
    $convcpu_i$ &amount of vCPU provisioned to $Con_i^j$ \\
    \hline
    $ET_i^j$ & execution time of $t_i$ in $Con_i^j$ \\
    \hline
    $ST_i^j$ & start time of $t_i$ in $Con_i^j$ \\
    \hline
    $VM_i$ & a VM instance \\
    \hline
    $PM_i$ & a PM instance \\
    \hline
    $FT_i^j$ & finish time of $t_i$ in $Con_i^j$ \\
    \hline
    $FT_{end}$ & finish time of $t_{end}$ \\
    \hline
    $FT_{last}^k$ & finish time of the last tasks executed in $VM_k$ \\
    \hline
    $ST_{first}^k$ & start time of the first task executed in $VM_k$ \\
    \hline
    $RT_r$ & response time of a user request $req_r$ \\
    \hline
    $price_k$ & hourly price of a VM instance $VM_k$ \\
    \hline
    $Cost_k$ & total cost of renting a VM instance $VM_k$ \\
    \hline
    $ART(T)$ & average response time over a time period $T$ \\
    \hline
    $Cost(T)$ & total cost over a time period of $T$ \\
    \hline
    $ACT_{VM}(T)$ & index set of active VM instances over a time period $T$ \\
    \hline
    $REQ(T)$ & index set of user requests over a time period $T$ \\
    \hline
    $num$ & number of requests \\
    \hline
    $budget(T)$ & cost budget of an application provider over $T$ \\
    \hline
    \end{tabular}
\end{center}
\end{table}

\begin{figure}[htb!]
    \centering
    \includegraphics[width=1\linewidth]{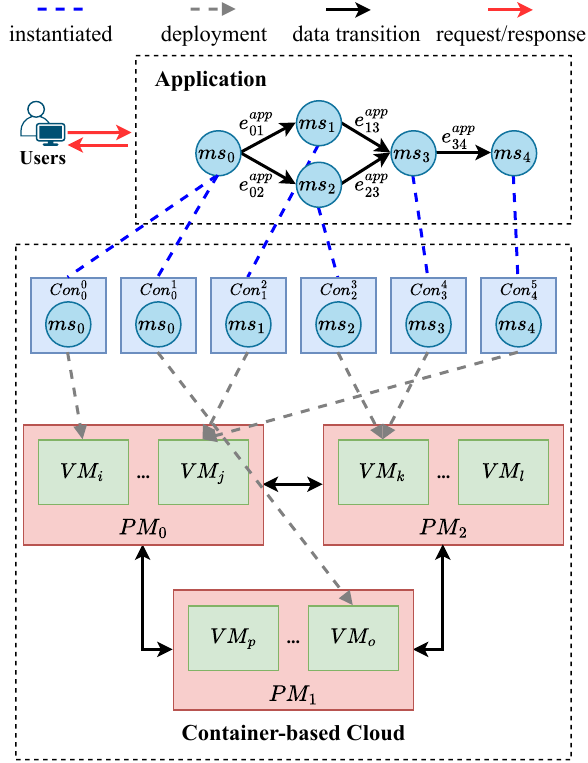}
    \caption{Microservice application deployed in the container-based cloud}
    \label{fig:problem}
\end{figure}

Fig.~\ref{fig:problem} presents the system model of the container-based cloud. A microservice application can be modeled as a Directed Acyclic Graph (DAG) $App=\langle V_{app}, E_{app}\rangle$, as shown in the application in Fig.~\ref{fig:problem}. $V_{app}=\{ms_0, ms_1, \dots, ms_n\}$ represents $n$ microservices. $e^{app}_{ij} \in E_{app}$ denotes the execution dependency between a pair of adjacent microservices $ms_i$ and $ms_j$. Following existing studies~\cite{wang2020elastic,bao2019performance,fang2023energy}, a microservice $ms_i$ is instantiated by at least one container $Con_i^j$, where $j$ denotes the index of the container, as shown in Fig.~\ref{fig:problem}.

Each user request triggers the execution of a workflow instance $WF=\langle V_{wf}, E_{wf}\rangle$~\cite{wang2020elastic}. $V_{wf} = \{t_{start}, t_1, t_2, \dots, t_n, t_{end}\}$ denotes tasks in a user request, where $t_{start}$ and $t_{end}$ are dummy starting and ending tasks, respectively. $e^{wf}_{ij} \in E_{wf}$ represents $t_i$ is the predecessor task of $t_j$ while $t_j$ is the successor task of $t_i$. Task $t_i$ can only be executed by a container of the corresponding microservice $ms_i$.

Let $et_i$ denote the execution time of $t_i$ with one vCPU, and $concpu_i^j$ denotes the amount of vCPU provisioned to $Con_i^j$. This study focus on resource adjustment of vCPU~\cite{chouliaras2023adaptive,shi2023auto,park2024graph}. This is because existing studies showed that CPU is the dominant factor affecting microservice application response time~\cite{park2024graph,lee2024comparative,wen2025statuscale}. Accordingly, the execution time of $t_i$ in $Con_i^j$ is
\begin{equation}
    ET_i^j = \frac{et_i}{concpu_i^j}.
\end{equation}

As assumed in existing studies~\cite{wang2020elastic,shi2020location,yang2022dual,bao2019performance}, a container can execute at most one task at any time. Meanwhile, each container maintains a pending queue of task, following~\cite{shi2020location,shi2023auto}. Each task starts execution only after the preceding task in the queue has been completed. As a result, the finish time $FT$ of a task $t_i$ in $Con_i^j$ is calculated by:
\begin{equation}
    FT_i^j = ST_i^j + ET_i^j,
\end{equation}
where $ST_i^j$ indicates the start time of $t_i$ in $Con_i^j$. Particularly, Particularly, $ST_i^j$ is defined as
\begin{equation}
    ST_i^j = WT^j_i + FT^{pre_i},
\label{eq:start_time}
\end{equation}
where $WT^j_i$ is the waiting time of $t_i$ in the pending queue of $Con_i^j$. $FT^{pre_i}$ denotes the finish time of predecessor tasks ($t_{pre_i}$) of $t_i$.

Let $req_r$ represent a user request for a microservice application, the response time $RT_r$ of $req_r$ is calculated by:
\begin{equation}
    RT_r = FT_{end}
\end{equation}

As shown in Fig.~\ref{fig:problem}, each container is deployed in a VM instance, while each VM is deployed in a PM instance~\cite{wang2020elastic,tan2020cooperative,fang2023group,fang2023energy}. A VM/PM instance can host multiple container/VM instances. $VM_k = (vmcpu_k, price_k)$ represents a VM instance, where $vmcpu_k$ denotes the amount of vCPU provided by $VM_k$ (i.e., CPU capacity) and $price_k$ indicates the hourly rental fee. 
The CPU capacity of a PM instance constrains the total CPU capacity of the VMs deployed in it, which further limits the available amount of vCPU provisioned to containers deployed in those VMs. The rental fee $Cost_k$ of any VM instance $VM_k$ is calculated by:
\begin{equation}
    Cost_k = price_k \times \frac{FT_{last}^k - ST_{first}^k}{3600},
\end{equation}
where $FT_{last}^k$ and $ST_{first}^k$ are the finish time and the start time of the last task and the first task executed in $VM_k$, respectively. The total cost $Cost(T)$ of renting VMs over a period of time $T$ is calculate by:
\begin{equation}
    Cost(T) = \sum_{k \in ACT_{VM}(T)}{Cost_k},
\end{equation}
where $ACT_{VM}(T)$ is the index set of active VMs over $T$.

In this article, we evaluate the QoS of microservice applications by \emph{Average Response Time} ($ART$)~\cite{cheng2024geoscale,shi2023auto,bai2024drpc}. Therefore, the aim of the AMC problem is to autoscaling containers to minimize the $ART$ over a time period $T$ while the $Cost(T)$ is under a cost budget, which is formulated as
\begin{align}
    \min ART(T) = & \min \frac{\sum_{r \in REQ(T)}{RT_r}}{num} \\
    \text{s.t.} \quad & Cost(T) \leq budget(T)
\end{align}
where $REQ(T)$ is the index set of user requests over $T$ and $num$ is the number of requests. $budget(T)$ is the cost budget of an application provider given over $T$.

The number of user requests sent to microservice applications varies over time. Thus the autoscaling process needs to dynamically 1) identify the containers that require scaling, and 2) determine the optimal amount of scaling resources.

\begin{figure*}[htb!]
    \centering
    \includegraphics[width=0.9\linewidth]{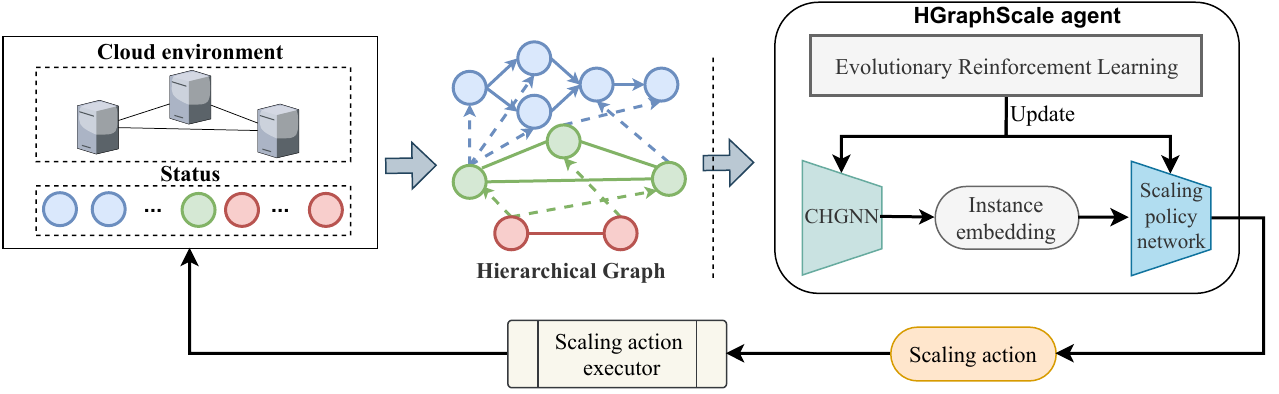}
    \caption{The overall framework of HGraphScale for the AMC problem.}
    \label{fig:HGraphScale}
\end{figure*}

\section{Proposed Autoscaling Approach}\label{section:HGraphScale}
\labeltext{The details of HGraphScale are introduced in this section.}\label{change:R2C8-3} \labeltext{Specifically, we model the process of the AMC problem as a Reinforcement Learning (RL) problem in Section~\ref{subsection:MDP}.}\label{change:R3C15-1} The overall framework of HGraphScale is shown in Fig.~\ref{fig:HGraphScale}. In each iteration, the state of the container-based cloud is extracted and represented as a hierarchical graph, detailed in Section~\ref{subsection:HGR}. \labeltext{Then, a novel \emph{Cloud-oriented Hierarchical Graph Neural Network (CHGNN)}}\label{change:R3C15-4} is proposed to learn the embedding of every container, introduced in Section~\ref{subsection:HGNN}. The learned container embedding is fed into a newly designed scaling policy network to produce scaling actions, as described in Section~\ref{subsection:policy}. A \emph{scaling action executor} performs either vertical or horizontal scaling in the container-based cloud based on the scaling actions, as outlined in Section~\ref{subsection:scaling}. Section~\ref{subsection:loadbalancer} presents how HGraphScale handles load balancing.

Evolutionary Reinforcement Learning (ERL)~\cite{huang2022cost,salimans2017evolution}, a widely recognized and practically popular algorithm, is leveraged to train the neural networks of HGraphScale. \labeltext{This is because ERL demonstrates strong exploration ability, ensures a stable training process, and requires relatively few hyperparameters for fine-tuning~\cite{huang2022cost,salimans2017evolution}. Moreover, recent studies have shown its effectiveness in several cloud-related applications~\cite{huang2022cost,shen2024cost}. The detailed process of training by ERL is provided in Section~\ref{subsection:es}.}\label{change:R3C2}



\subsection{RL Formulation}\label{subsection:MDP}
\labeltext{We formulate the process of solving AMC problems as an RL problem.}\label{change:R3C15-2} Particularly, at each decision step $t$, the cloud environment provides the state $s_t$ as a hierarchical graph. The HGraphScale agent in Figure~\ref{fig:HGraphScale} generates a scaling action $a_t$ based on $s_t$. 
The environment then performs $a_t$ and transitions to the next state $s_{t+1}$. 
\labeltext{The key components of this RL problem are outlined below.}\label{change:R2C3-9}

\subsubsection{State}
Each state $s_t$ is a snapshot of the status of the PMs, VMs and containers in the container-based cloud at a decision step $t$. We design a novel hierarchical graph $\mathcal{H}=\langle\mathcal{V}, \mathcal{E}\rangle$ to represent the states $s_t$, detailed in Section~\ref{subsection:HGR}.

The status of a PM instance $p$ is defined as $\vec{h_p^{pm}} = \{\mu_{pm_p}, \Omega_{pm_p}\}$, which denotes the resource utilization ($\mu_{pm_p}$) and the capacity ($\Omega_{pm_p}$) of $pm_p$, respectively. Status of a VM instance $v$ is defined as $\vec{h_v^{vm}} =\{\mu_{vm_v}, \Omega_{vm_v}, price_{vm_v}, rental_{vm_v}, art_{vm_v}\}$, which indicates the resource utilization ($\mu_{vm_v}$), the capacity ($\Omega_{vm_v}$), the per hour price ($price_{vm_v}$), the current rental fees ($rental_{vm_v}$) of $vm_v$, and the average response time ($art_{vm_v}$) of containers that are deployed in $vm_v$. 

Similarly, status of a container $c$ is defined as $\vec{h_c^{con}} = \{\Omega_{con_c}$, $\zeta_{con_c}$, $d_{con_c}$, $pending_{con_c}$, $art_{con_c}$, $predicted_{con_c}\}$. Specifically, $\Omega_{con_c}$ denotes the resource capacity of $con_c$. $\zeta_{con_c}$ represents the remaining resources of the VM that hosts $con_c$, indicating that container autoscaling is constrained by the resources of its hosting VM. The degree of $con_c$ in the graph is denoted as $d_{con_c}$. Moreover, $pending_{con_c}$ denotes pending requests, $art_{con_c}$ the average response time, and $predicted_{con_c}$ the future workload of $con_c$. 

This article follows~\cite{cheng2023proscale} to employ an effective and efficient workload predicting method, i.e., the SMA method, to predict the number of future requests $predicted_{con_i}$ for a container $con_i$. The predicted future workload is based on the information from the historical workload.

\subsubsection{Action}
A scaling action $a_t$ at time $t$ of \emph{HGrapScale} is represented as a 2-dimensional tuple: $\langle Ind, Scale \rangle$. \labeltext{$Ind \in [0, n] \cap \mathbb{Z}^+$ denotes the index of the container that requires scaling. Here, $n$ is the current number of containers, which changes dynamically over time.}\label{change:R2C3-13} $Scale \in [-m, +m] \cap \mathbb{Z}^+$ indicates the amount of resources for scaling. The sign of $Scale$ determines whether to increase or decrease the provisioned resources for container $Ind$. If $Scale$ equals $0$, it indicates that the resource provisioned to the $Ind$ container remains unchanged.

\subsubsection{Optimization objective}
To minimize the $ART(t)$ and ensure the cost adheres to the budget, the optimization objective of this RL problem is defined as
\begin{equation}
\label{eq:reward}
    Obj(T) = -ART(T)-\rho\cdot\max\big(0, (Cost(T)-budget(T))\big)
\end{equation}
where $Obj(T)$ is the objective value over a time period $T$ and $\rho$ controls the penalty intensity when the cost exceeds the given budget.

\begin{figure*}[htb!]
    \centering
    \resizebox{0.8\textwidth}{!}{
    \subfloat[Graph structures]{
        \includegraphics[width=0.8\textwidth]{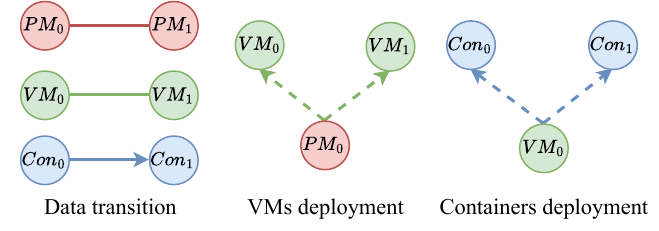}
        \label{subfig:graph_components}
    }
    \hfill
    \subfloat[Hierarchical graph]{
        \includegraphics[width=0.7\textwidth]{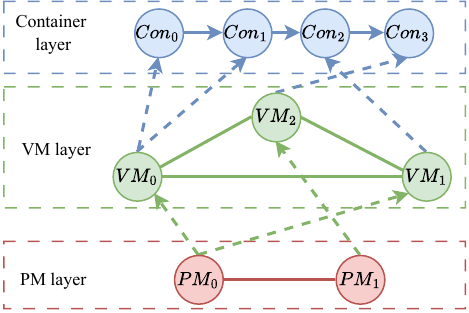}
        \label{subfig:hierarchical_graph}
    }
    }
    \caption{An example of the hierarchical graph representation of the container-cloud.}
    \label{fig:example_higraph}
\end{figure*}

\subsection{Hierarchical Graph Representation}\label{subsection:HGR}
We propose a hierarchical graph $\mathcal{H}=\langle\mathcal{V}, \mathcal{E}\rangle$ to represent the state of the container-based cloud, where node set $\mathcal{V} = \mathbb{C} \cup \mathbb{V} \cup \mathbb{P}$ consists of sets of container nodes $\mathbb{C}$, VM nodes $\mathbb{V}$ and PM nodes $\mathbb{P}$. Edge set $\mathcal{E} = \mathbb{E}_{depvm} \cup \mathbb{E}_{depcon} \cup \mathbb{E}_{pm} \cup \mathbb{E}_{vm} \cup \mathbb{E}_{con}$, where $\mathbb{E}_{depvm}$ and $\mathbb{E}_{depcon}$ represent the deployment scheme, $\mathbb{E}_{pm}$, $\mathbb{E}_{vm}$ and $\mathbb{E}_{con}$ represent the data transition between machines.

In particular, Fig.~\ref{fig:example_higraph}~\subref{subfig:graph_components} presents three main structures of a hierarchical graph. \labeltext{The edge between two PM nodes $PM_0, PM_1 \in \mathbb{P}$ is undirected $\{PM_0, PM_1\} \in \mathbb{E}_{pm}$. It represents data transmission caused by interactions between containers running on the two PMs.}\label{change:R2C3-10} Similarly, the edge between two VM nodes $VM_0, VM_1 \in \mathbb{V}$ is also undirected $\{VM_0, VM_1\} \in \mathbb{E}_{vm}$. \labeltext{For container nodes, there exist execution orders between connected containers. Thus, the edge between two container nodes $Con_0, Con_1 \in \mathbb{C}$ is directed, i.e., $(Con_0, Con_1) \in \mathbb{E}_{con}$.}\label{change:R2C3-11} Directed edges $(PM_0, VM_0), (PM_0, VM_1) \in \mathbb{E}_{depvm}$ in the \emph{VMs deployment} structures, indicating $VM_0$ and $VM_1$ are deployed in $PM_0$. Likewise, if $Con_0$ and $Con_1$ are deployed in $VM_0$, there are directed edges $(VM_0, Con_0), (VM_0, Con_1) \in \mathbb{E}_{depcon}$.

At each decision step $t$, the state $s_t$ of the container-based cloud is represented by the hierarchical graph, as shown in Fig.~\ref{fig:example_higraph}~\subref{subfig:hierarchical_graph}. The hierarchical graph consists of the \emph{PM layer}, \emph{VM layer} and \emph{Container layer}. The raw features of each node include the status of the corresponding machine (container, VM instance or PM instance), as described in Section~\ref{subsection:MDP}. \labeltext{We denote PM features as $\mathbf{h^{pm}} = \{\vec{h_0^{pm}}, \dots, \vec{h_P^{pm}}\}$, VM features as $\mathbf{h^{vm}} = \{\vec{h_0^{vm}}, \dots, \vec{h_V^{vm}}\}$, and container features as $\mathbf{h^{con}} = \{\vec{h_0^{con}}, \dots, \vec{h_C^{con}}\}$. The values $P$, $V$, and $C$ correspond to the numbers of PMs, VMs, and containers, respectively.}\label{change:R2C3-12} Our newly designed CHGNN learns container embedding from this hierarchical graph and the raw features of each node.

\subsection{Cloud-Oriented Hierarchical Graph Neural Network\label{subsection:HGNN}}
\begin{figure}[htb!]
    \centering
    \includegraphics[width=1.04\linewidth]{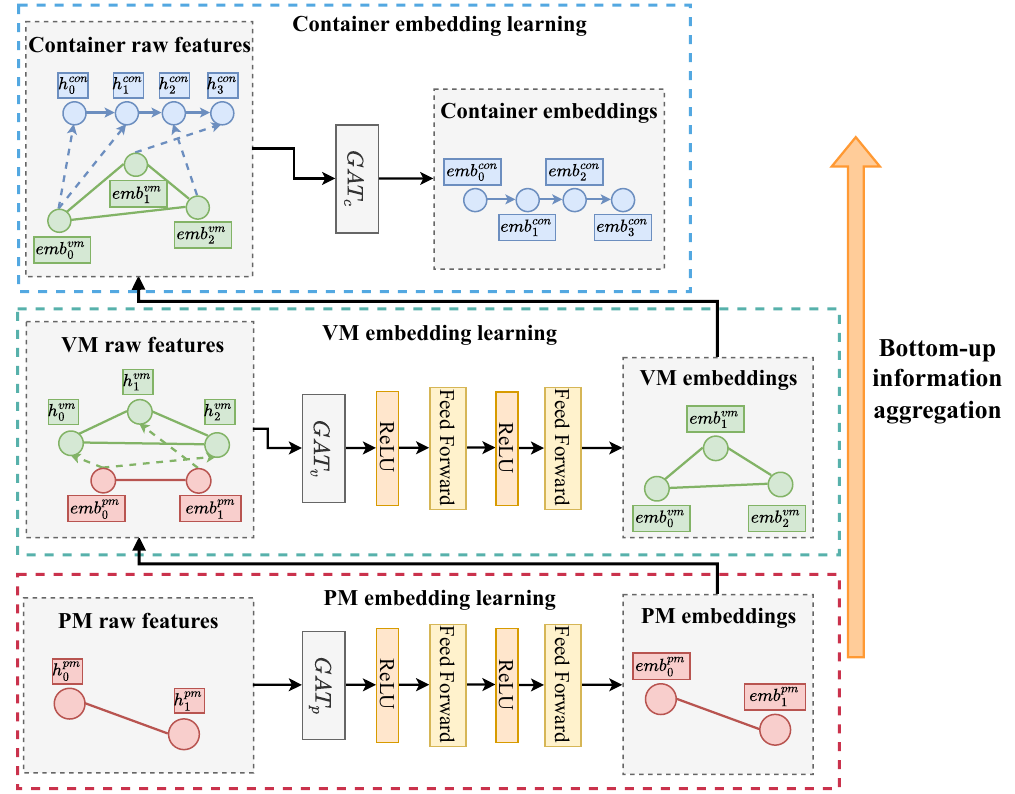}
    \caption{The architecture of the CHGNN.}
    \label{fig:CHGNN}
\end{figure}

Given the hierarchical graph represented state, we proposed CHGNN to learns container embedding progressively through a bottom-up information aggregation mechanism, as shown in Fig.~\ref{fig:CHGNN}. Specifically, HCGNN first learns PM embedding in the \emph{PM layer}, then PM embedding is propagated to the \emph{VM layers} for VM embedding learning. At last, VM embedding is propagated to the \emph{Container layer} for container embedding learning. Details of embedding learning in each layer and the bottom-up information aggregation are provided as follows.

\subsubsection{Machine embedding Learning}
\labeltext{We stack \emph{graph attention layers}~\cite{velickovic2017graph} to construct Graph Attention Networks (GATs) (i.e., $GAT_p$, $GAT_v$, and $GAT_c$ in Fig.~\ref{fig:CHGNN}) to learn embeddings of PMs, VMs and containers.}\label{change:R2C3-14} \labeltext{The input to each graph attention layer consists of a graph and its node features. It then applies attention weights to aggregate neighbor information, resulting in updated node features.}\label{change:R2C3-15}

\labeltext{Consider an example of learning PM embeddings by $GAT_p$. The $GAT_p$ dynamically assigns attention weights $\alpha_{i,j}$ to PM $p_i$ and its neighbor $p_j$ in the PM layer, indicating the importance of $p_j$'s features to $p_i$~\cite{velickovic2017graph}. The $\alpha_{i,j}$ is calculated by}\label{change:R2C3-16}
\begin{equation}
\label{eq:attention}
    \alpha_{i,j} = \frac{\exp{(\mathrm{LeakyReLU}(\vec{a}^{T}[\mathbf{W_p}\vec{h_i^{pm}}\|\mathbf{W_p}\vec{h_j^{pm}}]))}}{\sum_{k \in \mathcal{N}_i}\exp{(\mathrm{LeakyReLU}(\vec{a}^{T}[\mathbf{W_p}\vec{h_i^{pm}}\|\mathbf{W_p}\vec{h_k^{pm}}]))}},
\end{equation}
where $\mathbf{W_p} \in \mathbb{R}^{F' \times F}$ is a learnable weight matrix in $GAT_p$, transforming the input features into high-level features. $\|$ indicates the concatenation operation. $\vec{a} \in \mathbb{R}^{2F'}$ is the learnable weight vector of a feedforward network. Following~\cite{velickovic2017graph}, $\mathrm{LeakyReLU}$ is applied for non-linear transformation. $\mathcal{N}_i$ is the neighbor nodes of $p_i$ in the PM layer.

The updated features $\vec{h_i^{pm'}}$ of a $p_i$ is generated by a graph attention layer according to
\begin{equation}
\label{eq:output}
    \vec{h_i^{pm'}} = \boldsymbol{\sigma}(\sum_{j \in \mathcal{N}_i}\alpha_{i,j}\mathbf{W_p}\vec{h_i^{pm}}),
\end{equation}
where $\boldsymbol{\sigma}$ is the sigmoid activation function that enables modeling of nonlinearity. The $GAT_p$ outputs node embedding $\mathbf{emb^{pm}} = \{emb^{pm}_0, \dots, emb^{pm}_P\}$ of the PM layer after passing through multiple stacked \emph{graph attention layers}. The VM layer and container layer follow the same process of embedding learning by $GAT_v$ and $GAT_c$, respectively.

\subsubsection{Bottom-Up Information Aggregation}
In our proposed bottom-up information aggregation mechanism, the learned PM embedding $\mathbf{emb^{pm}} = \{emb^{pm}_0, \dots, emb^{pm}_P\}$ propagates to the \emph{VM layer}. Thus, the inputs of $GAT_v$ are the concatenation of VM raw features and PM embedding $\mathbf{h^{vm}}\|\mathbf{emb^{pm}}$. The VM embedding $\mathbf{emb^{vm}} = \{emb^{vm}_0, \dots, emb^{vm}_V\}$ are learned by the $GAT_v$ and feed forward networks.

Similarly, the inputs of the $GAT_c$ are the concatenation of container raw features and VM embedding $\mathbf{h^{con}}\|\mathbf{emb^{vm}}$, which outputs the container embedding $\mathbf{emb^{con}} = \{emb^{con}_0, \dots, emb^{con}_C\}$. \labeltext{Through bottom-up information aggregation, the \emph{Container layer} effectively incorporates global information of the container-based cloud into container embeddings. These embeddings allow the proposed \emph{scaling policy network} to make system-aware scaling decisions.}\label{change:R2C3-20}

\subsection{Scaling Policy Network}\label{subsection:policy}
\begin{figure}[htb]
    \centering    \includegraphics[width=0.8\linewidth]{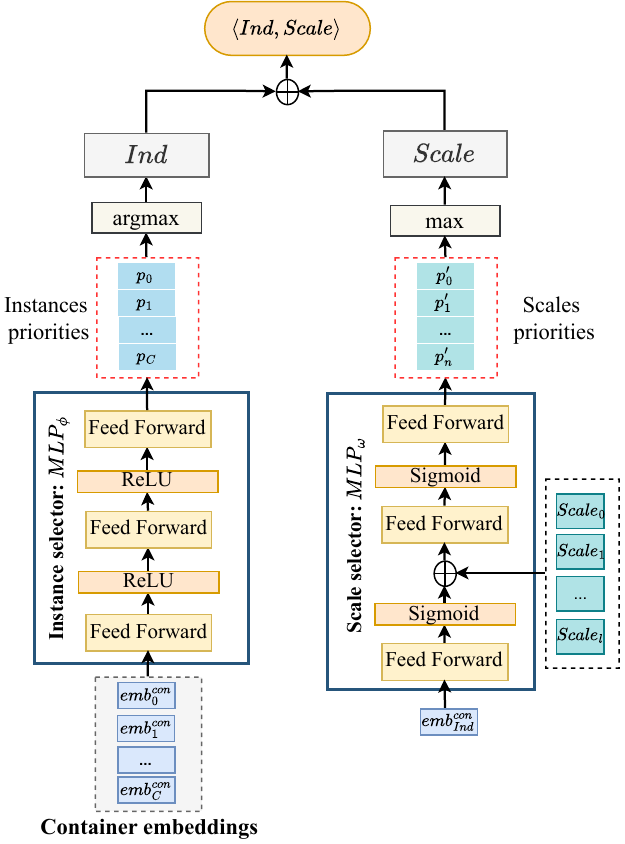}
    \caption{The architecture of policy network}
    \label{fig:policy}
\end{figure}

To generate a scaling action, we design a \emph{scaling policy network}, which takes the container embedding $\mathbf{emb^{con}} = \{emb^{con}_1, emb^{con}_2, \dots, emb^{con}_C\}$ as input and outputs scaling actions, as illustrated in Fig.~\ref{fig:policy}. \labeltext{A scaling action is defined as a tuple $\langle Ind, Scale \rangle$. To generate such actions, we design the scaling policy network with two MLPs: the \emph{instance selector} $MLP_{\phi}$ and the \emph{scale selector} $MLP_{\omega}$.}\label{change:R2C3-21}

$MLP_{\phi}$ is designed to calculate the priority values of each container $i$:
\begin{equation}
    p_i = MLP_{\phi}(emb^{con}_i).
\end{equation}
A container with a higher priority value implies a greater need for scaling. Thus, the index of the container to be scaled (e.g., $Ind$) is identified by
\begin{equation}
    Ind = \mathop{\arg\max}\limits_{i = 1, 2, \dots, C} \big(p_i\big),
\end{equation}
where $C$ is the current number of containers.

After identifying the container for scaling, the corresponding container embedding $emb^{con}_{Ind} \in \mathcal{R}^{1 \times d}$ is selected and fed into $MLP_{\omega}$. \labeltext{As shown in Fig.~\ref{fig:policy}, $emb^{con}_{Ind}$ is passed through a feedforward network. The output is then concatenated with a vector $\mathcal{S}  = \{S_0, S_1, \dots, S_l\}$, where $S_j \in \mathbb{Z}$ indicates the amount of scaling resource, resulting in a new vector $\mathcal{I} = \{emb^{con}_{Ind}||S_0, emb^{con}_{Ind}||S_1, \dots, emb^{con}_{Ind}||S_l\}$.}\label{change:R3C20} $\mathcal{I}$ is further processed by feed-forward networks. Finally, $MLP_{\omega}$ outputs the priority $p'_j$ of each $Scale_j \in \mathcal{S}$. Therefore, the amount of scaling resources is determined by
\begin{equation}
    Scale = \max_{j=0,1,2, \dots, l} (p_j).
\end{equation}

Afterwards, $Ind$ and $Scale$ are combined to create a complete scaling action $\langle Ind, Scale \rangle$.

\begin{algorithm}[htb!]
\begin{algorithmic}[1]
    \Require
        Scaling action: $\langle Ind, Scale \rangle$
    \Ensure vertical scaling or horizontal scaling
    \State $tar\_con \leftarrow container\_list[Ind]$ 
    \State $tar\_vm \leftarrow$ the VM that hosts $tar\_con$
    \If{$Scale > 0$}
    \State $max\_vcpu \leftarrow$ the remaining vCPUs of $tar\_vm$
        \If{$max\_vcpu > Scale$} \Comment{vertical scaling}
        \State Increase $Scale$ vCPUs to $tar\_con$
        \Else \Comment{horizontal scaling}
        \State Increase $max\_vcpu$ vCPUs to $tar\_con$
        \State $vcpu \leftarrow Scale - max\_vcpu$
        \State Create a new container with $vcpu$ vCPUs
        \EndIf
    \Else
    \State $con\_vcpu \leftarrow$ number of vCPUs provisioned to $tar\_vm$ 
        \If{$con\_vcpu > Scale$} \Comment{vertical scaling}
        \State Decrease $Scale$ vCPUs to $tar\_con$
        \Else \Comment{horizontal scaling}
        \State Delete $tar\_con$
        \EndIf
    \EndIf
    
\end{algorithmic}
\caption{Scaling action executor}
\label{algorithm:scaling}
\end{algorithm}

\subsection{Scaling Action Executor}\label{subsection:scaling}
The scaling action executor transforms the scaling action $\langle Ind, Scale \rangle$ to vertical scaling, horizontal scaling or both. Algorithm~\ref{algorithm:scaling} summarizes the process of scaling action executor. Firstly, a container $tar\_con$ is selected based on $Ind$ (line 1). 

If $Scale > 0$, the scaling action executor increases the resource provisioned to $tar\_con$ (lines 3 to 12). To be specific, if the remaining CPU capacity $max\_vcpu$ of the VM hosting the $tar\_con$ is larger than the $Scale$, vertical scaling is applied to provision $Scale$ vCPUs to the container (line 6); otherwise, $max\_vcpu$ vCPUs are provisioned to $tar\_con$, then a new container is created by horizontal scaling. The newly created container is provisioned $(Scale - max\_vcpu)$ vCPUs (lines 7 to 11). 

The scaling action executor reduces the resource provisioned to $tar\_con$ when $Scale<0$ (lines 12 to 19). If $Scale$ is larger than the total vCPUs of $tar\_con$, the vCPUs of $tar\_con$ are reduced by the $Scale$ number (line 15). Otherwise, the container $tar\_con$ is deleted as a result of horizontal scaling (line 17). Note that the scaling action executor allows a microservice to be encapsulated within containers with heterogeneous resources, which can reduce resource wastage~\cite{wen2024combofunc,srirama2020application}. As a result, a load balancer is implemented to dispatch user requests among heterogeneous containers, as detailed in Section~\ref{subsection:loadbalancer}.

\subsection{Capacity-based Load Balancing}\label{subsection:loadbalancer}
\labeltext{HGraphScale applies Capacity-based Weighted Round-Robin (CWRR)~\cite{saidu2014load,huang2020scalable,shi2023auto,baresi2021kosmos} to dispatch user requests to a suitable container for the purpose of load balancing. Specifically, the weight $\mathcal{W}_j$ of a container $Con_i^j$ is determined by
\begin{equation}
    \mathcal{W}_i^j = \frac{\gamma_i^j}{\sum_{k \in set(ms_i)}\gamma_i^k},
\end{equation}
where $\gamma_i^j$ indicates the resource allocation of $Con_i^j$ and $set(ms_i)$ denotes the container set of microservice $ms_i$.}\label{change:R3C11-1}

\labeltext{The rationale for adopting CWRR in HGraphScale is threefold. First, CWRR is widely employed in practice owing to its simplicity~\cite{saidu2014load}. Second, CWRR demonstrates low computation overhead in handling load balancing. Third, it provides effective load balancing by dispatching more user requests to containers with higher capacities. Thus, CWRR can prevent any container from being heavily utilized, reducing long tail response times~\cite{huang2020scalable,baresi2021kosmos}.}\label{change:R3C11-2}

\subsection{Evolutionary Reinforcement Learning}\label{subsection:es}
\labeltext{In this article, we adapt ERL~\cite{salimans2017evolution} to train the neural networks of HGraphScale. ERL is a population-based approach to estimate the gradients of neural networks. Algorithm~\ref{algorithm:esrl} presents the pseudo-code of the ERL.}\label{change:R3C1-1}

\begin{algorithm}[htb!]
\begin{algorithmic}[1]
    \Require
        Population size: $N$, maximum generation: $max\_gen$, initial policy parameters: $\hat{\theta}$, learning rate: $\eta$, multi-variance gaussian noise standard deviation: $\sigma$
    \Ensure
        Trained neural network
    \State $gen \leftarrow 0$
    \While{$gen \leq max\_gen$}
        \For{$i = 0$ to N}
        \State Sample perturbation $\epsilon_i \sim \mathcal{N}(0, 1)$
        \State Update the neural network by using $\theta_i \leftarrow \hat{\theta} + \sigma\epsilon_i$
        \State Calculate Fitness $F(\theta_i)$ based on Eq.~\ref{eq:reward}
        \EndFor
        \State Estimate policy gradient $\nabla_{\theta}\mathbb{E}_{\epsilon_i \sim \mathcal{N}(0, 1)}F(\hat{\theta} + \sigma\epsilon_i)$
        \State $\hat{\theta} \leftarrow \hat{\theta} + \sigma F(\hat{\theta} + \sigma\epsilon_i)$
    \EndWhile
    
\end{algorithmic}

\caption{Evolutionary Reinforcement Learning (ERL)}
\label{algorithm:esrl}
\end{algorithm}

\labeltext{In particular, the CHGNN and scaling policy network of HGraphScale initial all trainable parameters $\hat{\theta} = \{\vec{a}, W_p, W_v, W_c, \phi, \omega\}$, randomly. Each iteration starts with sampling $N$ perturbations $[\epsilon_i]_{i = 0, 1, \ldots, N}$ from standard gaussian distribution $\mathcal{N}(0, 1)$ (line 4). Then, a population of $N$ individuals $[\theta_i]_{i = 0, 1, \ldots, N}$ is generated by adding noise to $\hat{\theta}$ (line 5).}\label{change:R3C1-2}

\labeltext{The fitness of an individual $\theta_i$ is evaluated based on the optimization objectives defined in Eq.~\eqref{eq:reward} (line 6), which is calculated by
\begin{equation}
    F(\theta_i) = Obj(T).
\end{equation}}\label{change:R3C1-3}

\labeltext{Then, the parameters of the policy network are updated by the estimated gradient, which is the expectation of individuals' fitness (line 8). Specifically, the gradient is estimated by 
\begin{equation}
\begin{split}
    \nabla_{\theta}\mathbb{E}_{\epsilon \sim \mathcal{N}(0, 1)}F(\hat{\theta} + \sigma\epsilon) &= \frac{1}{\sigma}\nabla_{\theta}\mathbb{E}_{\epsilon \sim \mathcal{N}(0, 1)}[F(\hat{\theta} + \sigma\epsilon)\epsilon] \\ &\approx \frac{1}{N\sigma}\sum_{i=1}^N[F(\theta + \sigma\epsilon_i)\epsilon_i].
\end{split}
\end{equation}
Finally, the policy parameters are updated by gradient descent (line 9).}\label{change:R3C1-4}

\section{Performance Evaluation}\label{section:experiment}
\labeltext{In this section, we conduct comprehensive experiments to test the performance of our proposed HGraphScale.}\label{change:R2C8-4} We first present the setup of experiments, the HGraphScale configuration and the competing approaches. Then, the experiment results are shown. \labeltext{Code of implementation, dataset and configuration are made publicly available\footnote{https://github.com/sine-fandel/HGraphScale}.}

\subsection{Experiment Setup}\label{subsection:setup}
\labeltext{All experiments are carried out in a simulator that implemented based on OpenAI Gymnasium~\cite{gymnasium_doc}. The simulator models dynamic resource allocation across containers, VMs, and PMs, and reproduces fluctuating workload. It also simulates autoscaling behaviors with transient effects.}\label{change:R3C10}. The worst-case scenarios analysis in Appendix~B enhances the fidelity of the simulator to real-world environments.

\begin{figure}[htb!]
    \centering
    \includegraphics[width=1\linewidth]{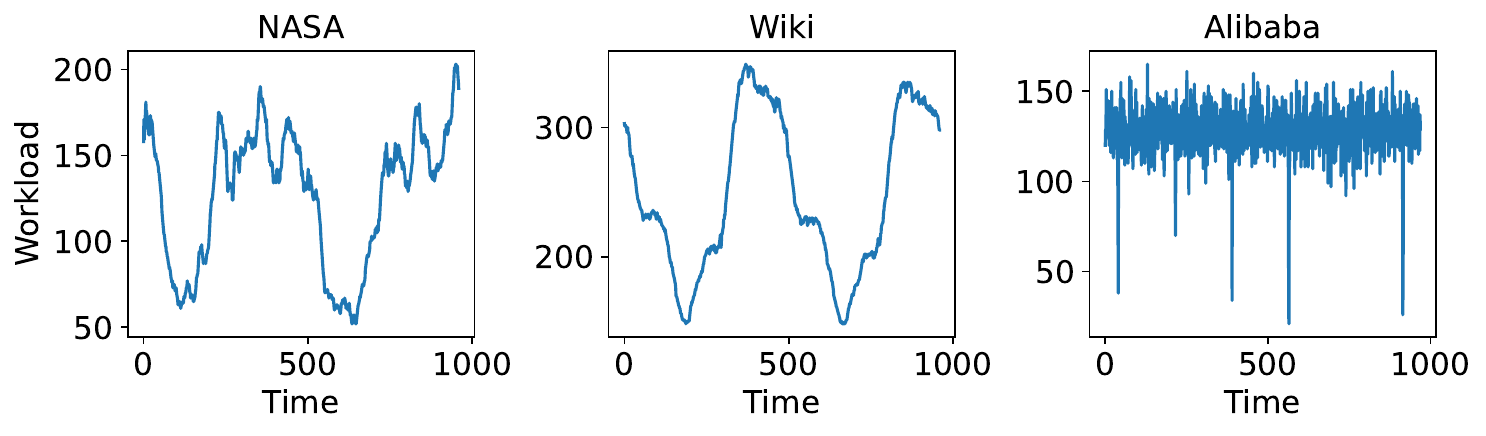}
    \caption{Traces of user requests.}
    \label{fig:workload}
\end{figure}

Three real-world traces of user requests, i.e., NASA\footnote{http://ita.ee.lbl.gov/html/traces.html}, Wiki\footnote{http://www.wikibench.eu/wp-content/uploads/2010/10/vanbaaren-thesis.pdf} and Alibaba\footnote{https://github.com/alibaba/clusterdata/tree/master/cluster-trace-microservices-v2021} are used to create workloads for our experiments. Fig.~\ref{fig:workload} illustrates the workload patterns over the 960-time-unit period (2 days) of NASA, Wiki and Alibaba, with each time unit representing a 3-minute interval. The workload trace patterns are shown in Fig.~\ref{fig:workload}. The first 480 time units (one day) of workload from NASA or Wiki are extracted for training, while the remaining time units of workload are used for test~\cite{shi2023auto}. In this article, a scaling action is made every 3 minutes, following~\cite{shi2023auto}.

\begin{figure}[htb!]
    \centering
    \includegraphics[width=0.9\linewidth]{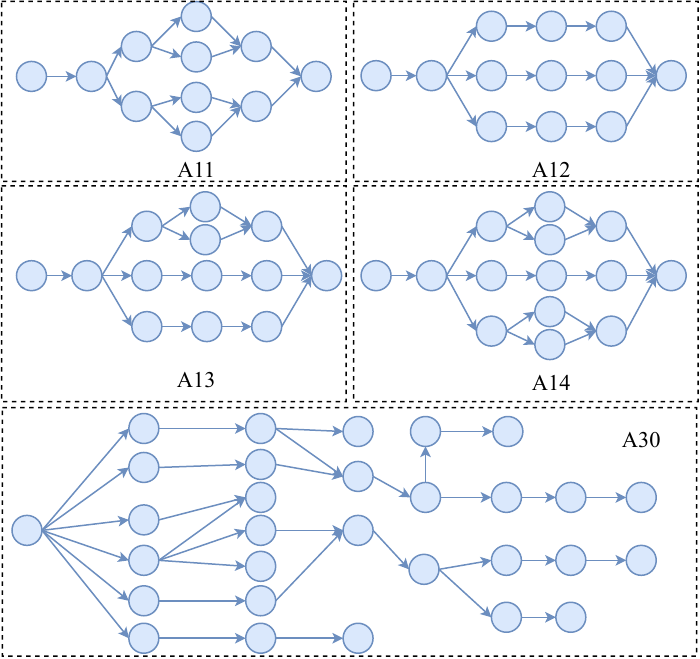}
    \caption{Microservice applications used in experiments}
    \label{fig:applications}
\end{figure}

Four medium-scale microservice applications~\cite{shi2020location,huang2015service,shi2021cost} and a large-scale microservice application~\cite{cheng2023proscale} are used for our experiments, as summarized in Fig.~\ref{fig:applications}. Each microservice application has a different number of microservices and application structures. For convenience, we denote them as ``A11'', ``A12'', ``A13'', ``A14'' and ``A30'', according to their microservices number. 

Moreover, the cloud environment is equipped with 5 VM types from Amazon EC2\footnote{https://aws.amazon.com/ec2/pricing/on-demand/}. The details of VM types are summarized in Tabel~\ref{table:vmtypes}. Each PM in the cloud environment has 64 vCPUs and 3200 GiB, following~\cite{tan2020cooperative,wang2023energy}.

\begin{table}[htb!]
\caption{Five VM types used in experiments}
\label{table:vmtypes}
\begin{center}
    \begin{tabular}{|c|c|c|c|}
    \hline
    \textbf{VM type} & \textbf{vCPU} & \textbf{Memory (GiB)} & \textbf{Hourly price (\$)} \\
    \hline
    m5.xlarge & 4 & 16 & 0.192 \\
    \hline
    m5.2xlarge & 8 & 32 & 0.384 \\
    \hline
    m5.4xlarge & 16 & 64 & 0.768 \\
    \hline
    m5.8xlarge & 32 & 128 & 
    1.536 \\
    \hline
    m5.12xlarge & 48 & 192 & 2.304  \\
    \hline
    \end{tabular}
\end{center}
\end{table}

To sum up, there are 15 scenarios designed for experiments based on three real-world traces and five types of microservice applications. \labeltext{In the initial stage of each scenario, each microservice is instantiated with a container, allocated with a vCPU and evenly deployed across three ``m5.4xlarge'' VMs~\cite{fang2025communication,tan2020cooperative}. This seting allows each VM has enough remaining resources to support further vertical scaling.}\label{change:R3C7}

\subsection{HGraphScale Configuration}
This article sets the number of \emph{graph attention layers} as: in container layer $L_c = 2$, in VM layer $L_v = 1$ and PM layer $L_p = 1$, respectively. The dimension of GAT's output is 64. The hidden dimension of each feedforward network is set as 64. 

All the hyperparameter settings of the ERL follow existing studies~\cite{huang2022cost} that are designed for practical application. Specifically, we set the population size of the ERL as 40. The maximum generation is set as 1000, while HGraphScale converges at about 400 generations in all scenarios. The learning rate $\eta$ and the Gaussian noise standard deviation of ERL $\sigma$ are set as 0.01 and 0.05, respectively. The parameters are updated by Adam Optimizer. The $budget(T)$ of the optimization objective Eq.~\ref{eq:reward} is set as 200 USD per day~\cite{cheng2024geoscale,xie2024pbscaler}, while the performances under different budgets are evaluated in Section~\ref{subsection:further}. The penalty $\rho$ is set as 100, following~\cite{shi2023auto}. The performances of HGraphScale under different penalty settings are discussed in Section~\ref{subsubsection:penalty}.

\subsection{Competing Approaches}\label{subsection:competing}
HGraphScale is compared to two heuristic-based autoscaling approaches, two state-of-the-art DRL-based autoscaling approaches and a GNN-based autoscaling approaches. \labeltext{All competing approaches and HGraphScale share the same initial placement of containers.} \labeltext{Moreover, they deploy newly created containers from horizontal scaling into suitable VMs using the Best-Fit heuristic~\cite{jangiti2020hybrid}. With this heuristic, each new container is placed on the VM with the least remaining capacity that can still satisfy its demand. This strategy improves VM utilization and reduces the overall cost.}\label{change:R3C3}

\textbf{AWS-Scale}~\cite{AWSAutoScaling2022} is a threshold-based autoscaling approach. Referring to~\cite{shi2023auto,nouri2019autonomic}, we set the upper threshold as 0.8 and the lower threshold as 0.6 for CPU utilization of each container. If the CPU utilization of a container exceeds the upper threshold, a replica of this container is created. Conversely, if the CPU utilization of a container falls below the lower threshold, the container is removed.

\textbf{ProScale}~\cite{cheng2023proscale} is a heuristic-based proactive autoscaling method that leverages the SMA to predict future request workloads of each container. The horizontal scaling is made according to the gap between the predicted future workload and the current request processing rates. 

\labeltext{\textbf{DeepScale}~\cite{shi2023auto} is an autoscaling approach based on DQN. Specifically, it uses a deep neural network to make high-level decisions, i.e., increase, decrease and maintain the amount of resources provisioned to containers. Then, heuristics based on queue theory is proposed to make low-level scaling actions, including horizontal scaling and vertical scaling.}\label{change:R3C5-1}

\labeltext{\textbf{DRPC}~\cite{bai2024drpc} is a distributed reinforcement learning approach for autoscaling. It first trains a central module using Twin Delayed Deep Deterministic Policy Gradient. After training the central module, multiple deployment units are trained to imitate the central module's behaviors. Deployment units make scaling actions (horizontal scaling and vertical scaling) for each microservice in a distributed manner.}\label{change:R3C5-2}

\labeltext{\textbf{AGQ}~\cite{liang2026autoscaling} applies Graph Convolution Network (GCN) for resource estimation. The predicted future resource demand is utilized to make horizontal scaling decisions, i.e., increase replicas, reduce replicas and no operation. The resource adjustment agent is trained by Q-learning.}\label{change:R3C4}

\begin{table*}[htb!]
\caption{Performance comparisons in terms of ART (ms) and the violation degree (``Vio''), which is defined as the percentage of cost exceeding the budget (200 USD).}
\label{table:result}
\centering
\begin{tabular}{c|cc|cc|cc|cc|cc|cc}
\hline
\multirow{2}{*}{Scenario} & \multicolumn{2}{c}{AWS-Scale} & \multicolumn{2}{c}{ProScale} & \multicolumn{2}{c}{DeepScale} & \multicolumn{2}{c}{DRPC} & \multicolumn{2}{c}{AGQ} & \multicolumn{2}{c}{HGraphScale} \\
& ART & Vio & ART & Vio & ART & Vio & ART & Vio & ART & Vio & ART & Vio \\
\hline
NASA-11 & 410.42 & 0.00\% & 305.57 & 52.02\% & 306.60 & 0.00\% & 289.92 & 0.00\% & 278.14 & 0.00\% & \textbf{255.12} & 0.00\% \\
NASA-12 & 688.52 & 0.00\% & 387.72 & 22.87\% & 532.65 & 0.00\% & 433.23 & 10.40\% & 538.63 & 13.39\% & \textbf{268.47} & 0.00\% \\
NASA-13 & 899.04 & 0.00\% & 406.82 & 0.82\% & 493.62 & 44.04\% & 532.34 & 0.00\% & 243.87 & 0.00\% & \textbf{178.34} & 0.00\% \\
NASA-14 & 1022.10 & 0.00\% & 532.34 & 28.37\% & 348.03 & 66.17\% & 510.72 & 1.66\% & 336.43 & 161.58\% & \textbf{325.67} & 0.00\%  \\
NASA-30 & 491.39 & 0.00\% & \textbf{303.49} & 70.11\% & 407.21 & 0.00\% & 391.94 & 0.00\% & 474.45 & 3.39\% & 389.46 & 0.00\%  \\
\hline
Wiki-11 & 489.73 & 0.00\% &  532.46 & 0.00\% & 318.29 & 34.74\% & 415.48 & 28.17\% & 361.74 & 41.01\% & \textbf{307.70} & 0.00\% \\
Wiki-12 & 864.65  & 0.00\% & 687.00 & 0.00\% & 549.98 & 0.00\% & 512.40 & 12.51\% & 457.01 & 36.43\% & \textbf{424.30} & 0.00\% \\
Wiki-13 & 1080.44 & 0.00\% & 482.13 & 13.18\% & 675.37 & 0.00\% & 491.68 & 56.17\% & \textbf{367.61} & 13.85\% & 369.16 & 0.00\% \\
Wiki-14 & 1022.10  & 0.00\% & 532.34 & 11.36\% & 348.03 & 26.62\% & 510.72 & 0.00\% & 520.24 & 11.27\% & \textbf{ 325.67} & 0.00\% \\
Wiki-30 & 395.68 & 0.00\% & 426.66 & 51.06\% & 388.31 & 56.10\% & 374.88 & 0.00\% & 488.24 & 10.28\% & \textbf{350.96} & 0.00\% \\
\hline
Alibaba-11 & 395.83 & 0.00\% & 476.67 & 0.00\% & 307.47 & 0.00\% & 249.67 & 0.00\% & 295.94 & 0.00\% & \textbf{222.17} & 0.00\%  \\
Alibaba-12 & 665.54 & 0.00\% & 654.04 & 0.00\% & 312.20 & 66.21\% & 291.86 & 0.00\% & 292.12 & 12.07\% & \textbf{283.78} & 0.00\% \\
Alibaba-13 & 525.74 & 0.00\% & 281.62 & 0.00\% & 212.43 & 55.71\% & 251.72 & 0.00\% & 237.93 & 0.00\% & \textbf{178.91} & 0.00\%  \\
Alibaba-14 & 988.76 & 0.00\% & 549.38 & 0.00\% & 327.00 & 20.28\% & \textbf{277.06} & 56.84\% & 421.87 & 12.71\% & 299.28 & 0.00\%  \\
Alibaba-30 & 474.71 & 0.00\% & 237.74 & 157.58\% & 210.13 & 12.78\% & 191.33 & 34.22\% & 442.87 & 0.00\% & \textbf{183.94} & 0.00\%  \\

\hline
\end{tabular}
\end{table*}



\subsection{Performance Comparison}\label{subsection:comparison}
\labeltext{TABLE~\ref{table:result} presents the test results on each scenario, where the best performance of ART in each scenario is highlighted in \textbf{bold}. Specifically, HGraphScale decreases from 37.17\% to 80.16\% of ART when compared to threshold-based AWS-Scale. This is because a fixed threshold setting cannot adapt effectively to workload changes across time. In the NASA-30 scenario, HGraphScale performs 28.32\% worse than ProScale in terms of ART. However, in this scenario, ProScale exceeds the budget by 70.11\%. In other 14 scenarios, HGraphScale achieved 16.51\% to 56.16\% less ART than ProScale.}\label{change:R1C1-1}

\labeltext{When compared to DeepScale, HGraphScale produces 3.33\% to 63.9\% less ART. As for DRPC, HGraphScale produces 7.42\% larger ART than DeepScale in Alibaba-14, while producing 3.33\% to 63.87\% less ART in the remaining scenarios. Although the ART of Alibaba-14 produced by DRPC is slightly better than HGraphScale, the corresponding cost exceeds the budget by 56.84\%. Although AGQ is also a GNN-based autoscaling method, it only shows a slight advantage over HGraphScale in Wiki-13. However, in this case, AGQ also exceeds the budget by 13.85\%.}\label{change:R1C1-2}

\labeltext{TABLE~\ref{table:result} also presents the violation degree (``Vio'') that quantifies the percentage of cost exceeding the predefined budget (200 USD). We can observe from this table that the total VM rental cost of HGraphScale is always kept under the budget. This indicates that HGraphScale can make suitable scaling actions to avoid resource wastage. In contrast, ProScale, DeepScale, DRPC and AGQ exceed the budget in multiple scenarios. Although AWS-Scale also prevents budget violation by removing containers promptly when their CPU utilization is under the lower threshold. However, this design makes AWS-Scale vulnerable to QoS degradation under dynamic workloads. Container removal during low request periods leads to increased ART when user demand rises abruptly.}\label{change:R1C1-3}



\subsection{Further Analysis}\label{subsection:further}

\subsubsection{Tail Response Time}\label{subsubsection:tail_response}

\begin{figure}[htb!]
    \centering
    \includegraphics[width=1\linewidth]{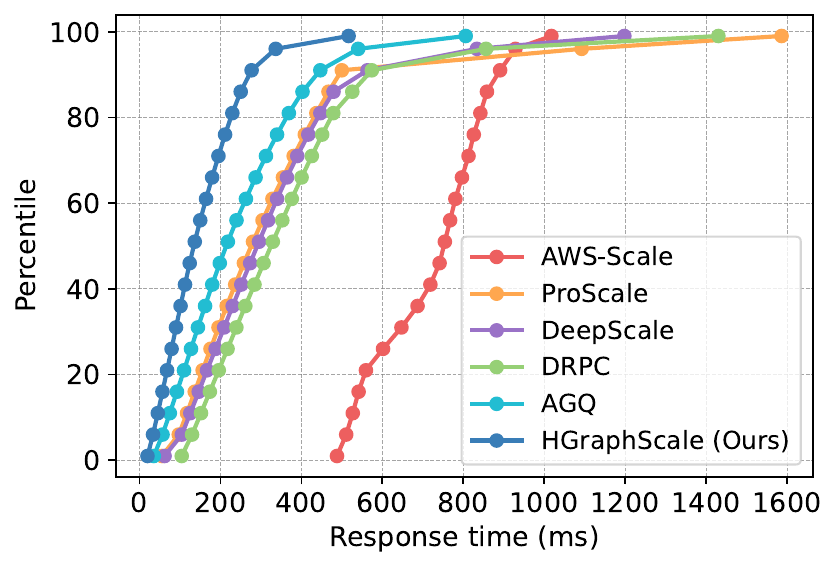}
    \caption{Response times of NASA-13 at different percentiles for AWS-Scale, ProScale, DeepScale, DRPC and HGraphScale.}
    \label{fig:percentile}
\end{figure}

\begin{figure}[htb!]
    \centering
    \includegraphics[width=1\linewidth]{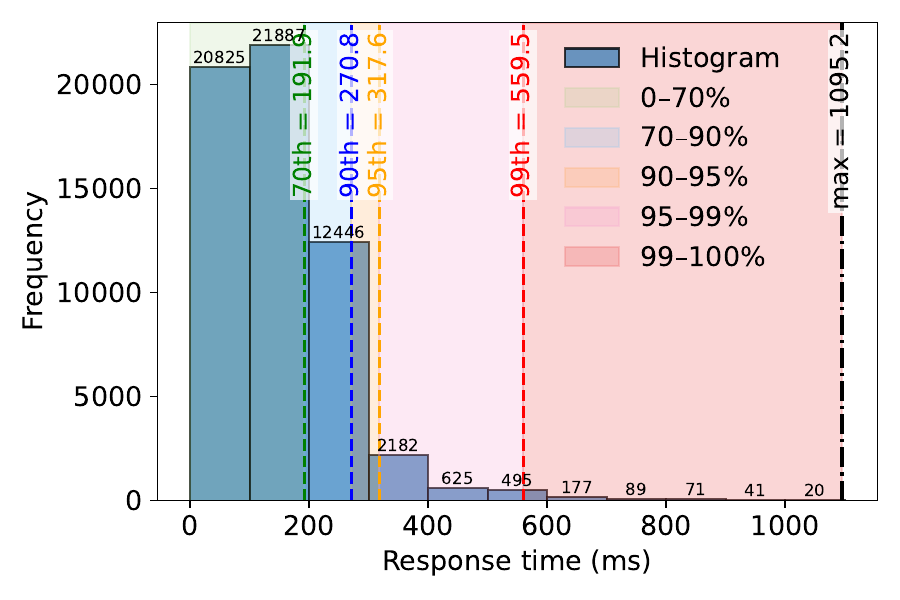}
    \caption{Response time distribution of HGraphScale in NASA-13}
    \label{fig:art_distribution}
\end{figure}

Besides the ART, the tail response time also provides insights into the QoS of microservice applications in the industry~\cite{dean2013tail,xie2018cutting,bai2024drpc,shi2023auto}. Fig.~\ref{fig:percentile} shows the maximum response times at different percentiles of user requests in NASA-13 (other scenarios have similar trends). We can see that HGraphScale achieves lower response times at all percentiles. Fig.~\ref{fig:art_distribution} provides the details response time distribution of HGraphScale in NASA-13. The results show that HGraphScale ensures 95\% of user requests are responded within 500 ms, and the maximum response time is 1.095s, showing stable performance and bounded worst-case latency. More details of response times analysis are provided in Appendix~A.

\subsubsection{Ablation Studies}\label{subsubsection:ablation}
\labeltext{To evaluate the effectiveness of the hierarchical graph learning, we conduct ablation studies by removing the PM layer of HGraphScale, giving rise to a variant named \emph{w/o PM}. Moreover, we design another variant of HGraphScale without both VM and PM layers, named \emph{w/o VM \& PM}. HGraphScale is compared with \emph{w/o PM} and \emph{w/o VM \& PM} on NASA-11, NASA-12, NASA-13 and NASA-14.}\label{change:R3C12-1}

\begin{figure}[htb!]
    \centering
    \includegraphics[width=1\linewidth]{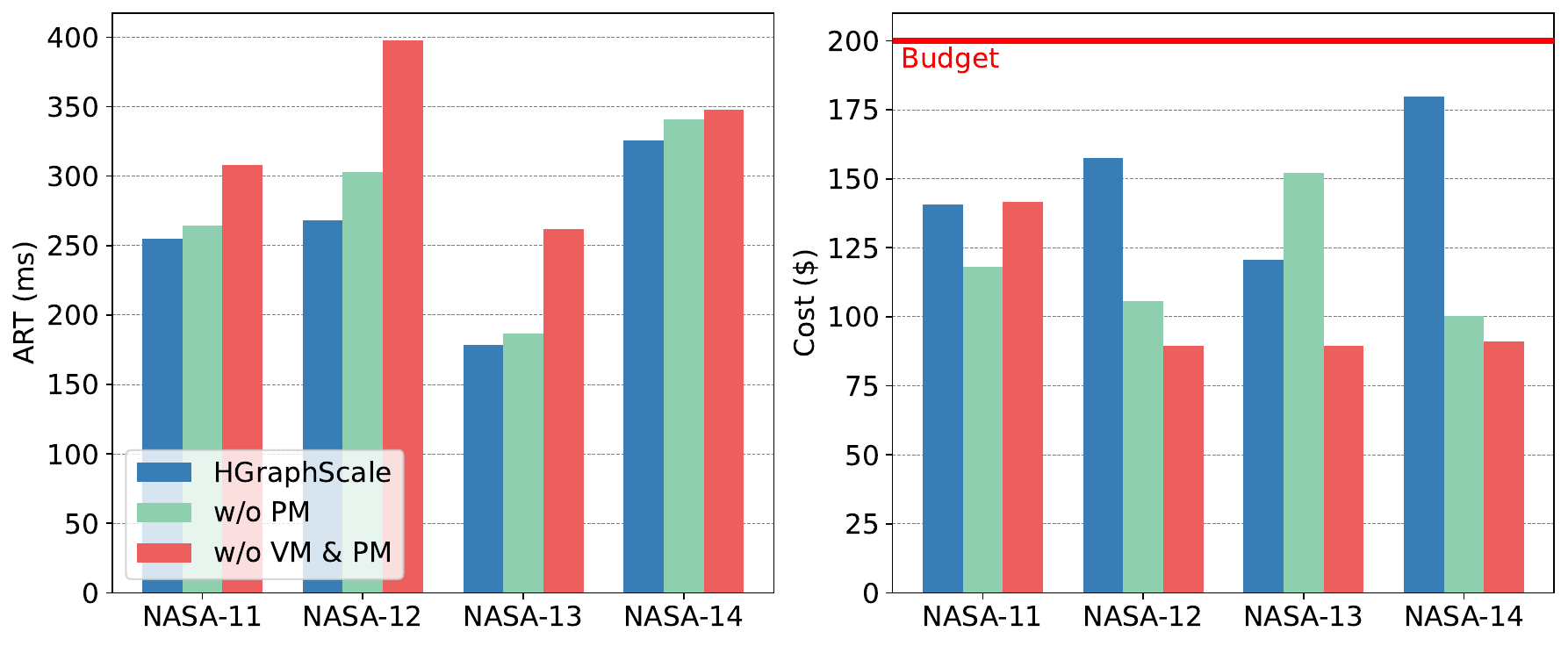}
    \caption{The comparison results of ablation studies under NASA workload}
    \label{fig:ablation}
\end{figure}

\labeltext{As shown in Fig.~\ref{fig:ablation}, both \emph{w/o PM} and \emph{w/o VM \& PM} ensure the cost is not exceed the budget. However, \emph{w/o PM} exhibits significantly inferior ART compared to HGraphScale, with \emph{w/o VM \& PM} performing even worse than \emph{w/o PM}. These results indicate the effectiveness of both VM and PM embedding learning in HGraphScale.}\label{change:R3C12-2} 

\begin{table}[htb!]
\centering
\caption{Performance Comparison With Different Budget: 150\$ and 250\$.}
\label{table:diffbudget}
\begin{tabular}{|c|c|c|c|c|}
\hline
\multirow{2}{*}{\textbf{Scenario}} & \multicolumn{2}{c|}{\textbf{150\$}} & \multicolumn{2}{c|}{\textbf{250\$}} \\ \cline{2-5} 
 & \textbf{ART (ms)} & \textbf{Cost (\$)} & \textbf{ART (ms)} & \textbf{Cost (\$)} \\ \hline
NASA-11 & 237.456 & 147.5719 & 219.7517 & 209.9346 \\
\hline
NASA-12 & 271.7395 & 91.1625 & 245.0896 & 209.3576 \\
\hline
NASA-13 & 430.8208 & 145.5375 & 162.9457 & 249.0529 \\
\hline
NASA-14 & 408.9306 & 140.1632 & 349.8992 & 247.2219 \\

\hline
\end{tabular}
\end{table}

\subsubsection{Performance Comparison with Different Budget}
We compare the performance of HGraphScale in solving the AMC problem with different cost budgets, that is, 150\$ and 250\$. TABLE~\ref{table:diffbudget} presents the ART and cost under different budgets. Specifically, both the stringent and relaxed budgets of the AMC problem can be satisfied by HGraphScale. We observe that the HGraphScale achieves lower ART under 250\$ budgets than under 150\$ budgets. The reason is that a relaxed budget allows for the provision of more resources to the containers.

\begin{figure}[htb!]
    \centering
    \includegraphics[width=1\linewidth]{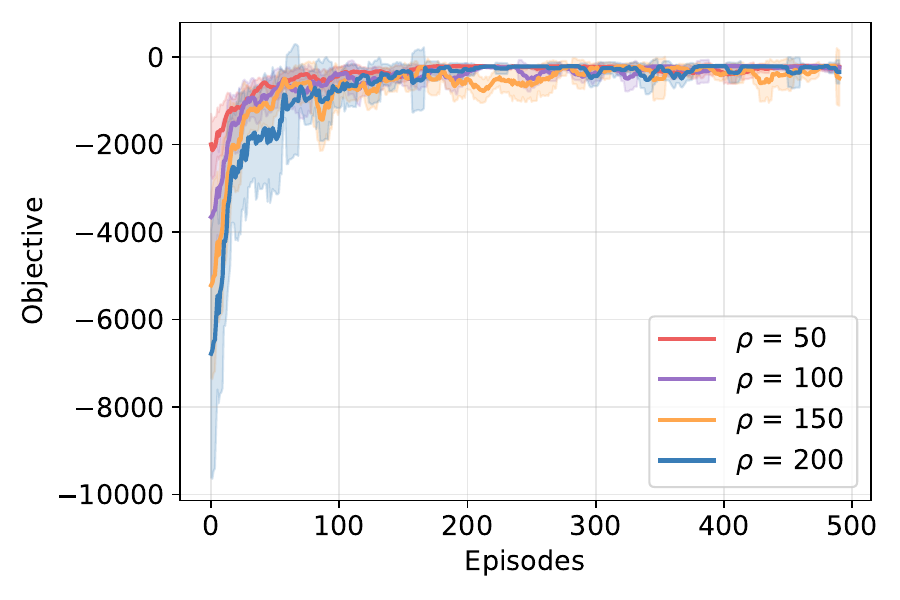}
    \caption{Training curve of HGraphScale under different settings of penalty $\rho$ on NASA-13}
    \label{fig:training_curve}
\end{figure}

\subsubsection{Performance Comparison with Different Penalty}\label{subsubsection:penalty}
\labeltext{The optimization objective of HGraphScale includes a penalty term ($\rho$) for violating the budget. Therefore, we conduct sensitivity analysis on different penalty settings, that is $\rho=50$, $\rho=100$, $\rho=150$ and $\rho=200$. Fig.~\ref{fig:training_curve} illustrates the training curves on NASA-13 obtained by different penalty settings. We can observe from this figure that the training process of HGraphScale is robust to different penalty settings, as they all achieve similar convergence stability.
}\label{change:R2C2-1}

\labeltext{TABLE~\ref{table:diffpenalty} presents the test performance of HGraphScale. HGraphScale ensures the cost under 200\$/day with different settings of $\rho$. Moreover, when $\rho = 50$, $\rho = 100$, and $\rho = 150$, HGraphScale achieves similar performances in terms of ART, while performance degradation occurs with $\rho = 200$. This is because the harsh penalty limits the exploration ability of HGraphScale during training.
}\label{change:R2C2-2}

\begin{table}[htb!]
\centering
\caption{Performance Comparison With Penalty Coefficient ($\rho$).}
\label{table:diffpenalty}
\begin{tabular}{|c|c|c|c|c|}
\hline
\multirow{2}{*}{\textbf{$\rho$}} & \multicolumn{2}{c|}{NASA-11} & \multicolumn{2}{c|}{NASA-13} \\ \cline{2-5} 
 & ART (ms) & Cost (\$) & ART (ms) & Cost (\$) \\ \hline
50 & 273.5127 & 145.5319 & 169.4707 & 133.7046 \\
\hline
100 & 255.1265 & 140.6622 & 178.3496 & 120.6937 \\
\hline
150 & 256.4128 & 117.3521 & 180.1922 & 146.2297 \\
\hline
200 & 324.0706 & 148.2200 & 203.6212 & 157.0951 \\

\hline
\end{tabular}
\end{table}

\subsubsection{Quantitative Analysis of Scaling Actions}\label{subsubsection:breakdown}
\begin{figure}[htb!]
    \centering
    \includegraphics[width=1\linewidth]{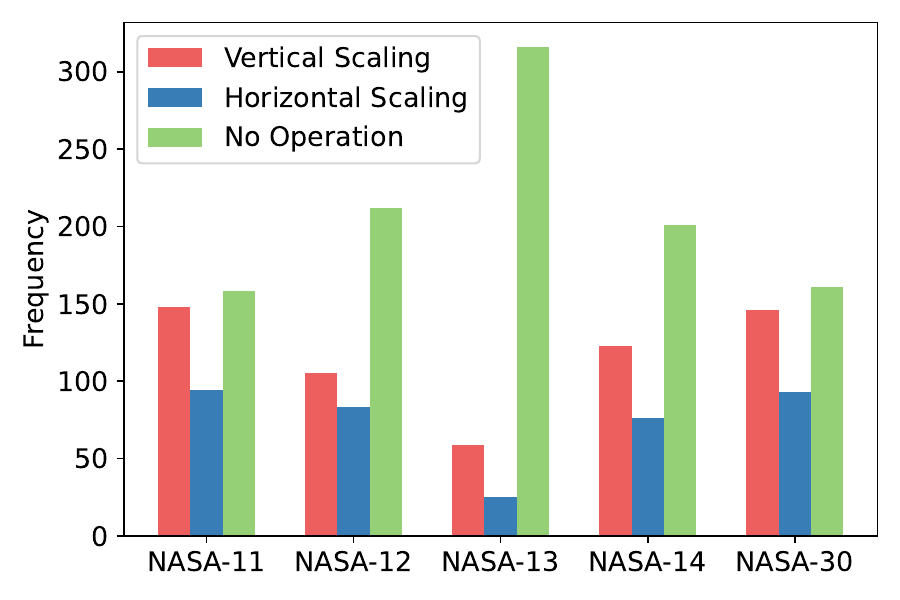}
    \caption{Quantitative breakdown of HGraphScale's scaling actions.}
    \label{fig:scaling_actions}
\end{figure}

\labeltext{To better understand the behavior of HGraphScale, we conduct a detailed analysis
of quantitative breakdown of autoscaling actions. Fig.~\ref{fig:scaling_actions} demonstrates the frequencies of scaling actions generated by HGraphScale, including \emph{vertical scaling}, \emph{horizontal scaling}, and \emph{no operation}. This figure provides evidences that HGraphScale tends to perform more vertical scaling than horizontal scaling in each scenario, resulting in fewer container replicas.}\label{change:R3C8-1}

\labeltext{Moreover, Fig.~\ref{fig:scaling_actions} also shows that \emph{no operation} dominates in all scenarios. These results indicate that HGraphScale improves application performance while maintaining system stability without frequent scaling. It further demonstrates HGraphScale’s ability to accurately identify containers requiring scaling and to determine appropriate scaling levels, thereby avoiding resource wastage.}\label{change:R3C8-2}



\section{Conclusion and Future Work}\label{section:conclusion}
In this article, we propose HGraphScale, a novel DRL-based autoscaling approach for microservice applications in container-based cloud. Particularly, We propose a hierarchical graph to capture dependencies in container-based clouds, a CHGNN with bottom-up aggregation to learn container embeddings, and a scaling policy network that makes scaling decisions based on these embeddings. The experimental results indicate that HGraphScale reduces average response time compared to threshold-based, DRL-based, and graph-based autoscaling, without exceeding the cost budget. In future work, we will investigate multi-resource autoscaling to further enhance our method. 

\bibliographystyle{splncs04}
\bibliography{bibliography}

\end{document}


\title{Appendix - HGraphScale: Hierarchical Graph Learning for Autoscaling Microservice Applications in Container-based Cloud Computing}

\author{Zhengxin Fang,~\IEEEmembership{Graduate Student Member,~IEEE,} Hui Ma,~\IEEEmembership{Senior Member,~IEEE,} Gang Chen,~\IEEEmembership{Senior Member,~IEEE,} and Rajkumar Buyya,~\IEEEmembership{Fellow,~IEEE}
}



\maketitle


\section{Response Time Distribution of HGraphScale}
\begin{figure*}[htb!]
    \centering
    \includegraphics[width=1\linewidth]{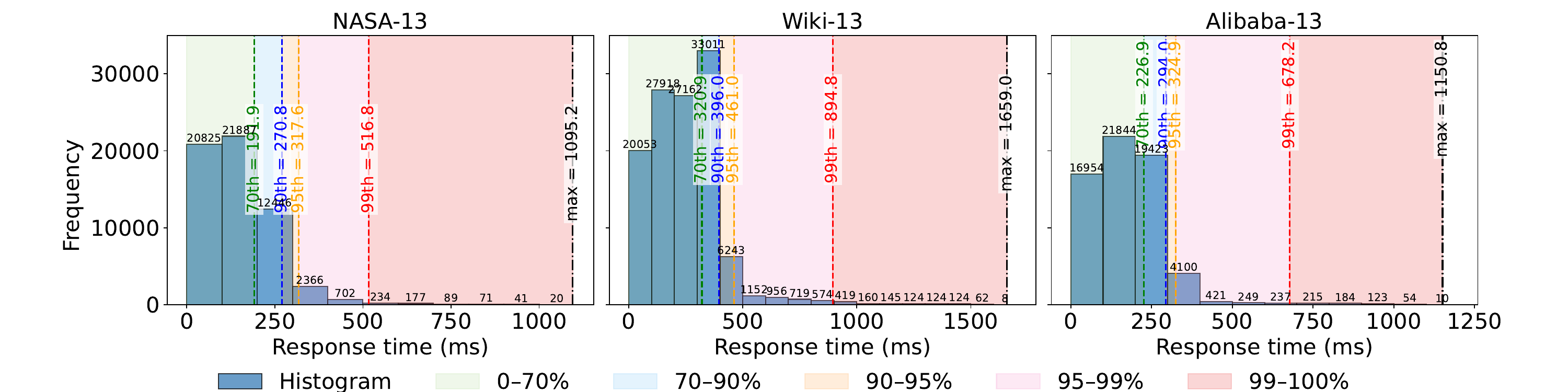}
    \caption{Distribution of response time (ms) in NASA-13, Wiki-13 and Alibaba-13}
    \label{fig:art_distribution}
\end{figure*}

With heterogeneous containers for the same service, the distribution of response times is skewed toward lower values, mainly because HGraphScale creates fewer containers. We conduct analysis on the distribution of ART, and the results indicate that HGraphScale performs well in terms of tail response times. Specifically, the ART distribution of HGraphScale is shown in Fig.~\ref{fig:art_distribution}. This figure presents the histogram and percentile of response times generated by HGraphScale in different scenarios. We can observe that 95\% of user requests are responded within 500 ms across all three traces, indicating the system delivers stable performance for most requests. Moreover, the maximum response time among these scenarios is 1.659 second, demonstrating that the system maintains bounded worst-case latency, thereby ensuring acceptable user experience even under extreme conditions.

\section{Worst-Case Scenarios Analysis}
The simulator we used is already considering the transient effect of scaling actions on response time. To further assess the influence of transient effects on the experiment, we examine worst-case scenarios. Specifically, in our worst-case setting, we assume that scaling operations incur the maximum transient overhead. For horizontal scaling, prior studies~\cite{park2024graph,baresi2021kosmos} and practical reports\footnote{https://www.cloudpilot.ai/en/blog/k8s-in-place-pod-resizing/}\footnote{https://cloud.google.com/kubernetes-engine/docs/how-to/monitor-startup-latency-metrics}
 show that creating a new replica may take up to 3 minutes before becoming fully operational; thus, our simulator sets the delay for horizontal scaling to 3 minutes. For vertical scaling, recent in-place scaling mechanisms\footnote{https://www.cloudpilot.ai/en/blog/k8s-in-place-pod-resizing/}\footnote{https://kubernetes.io/blog/2025/05/16/kubernetes-v1-33-in-place-pod-resize-beta/}\footnote{https://scaleops.com/blog/kubernetes-in-place-pod-vertical-scaling/}
 enable resource adjustments to running containers within microseconds to seconds; therefore, we conservatively set the worst-case delay to 10 seconds.

\begin{figure*}[htb!]
    \centering
    \includegraphics[width=1\linewidth]{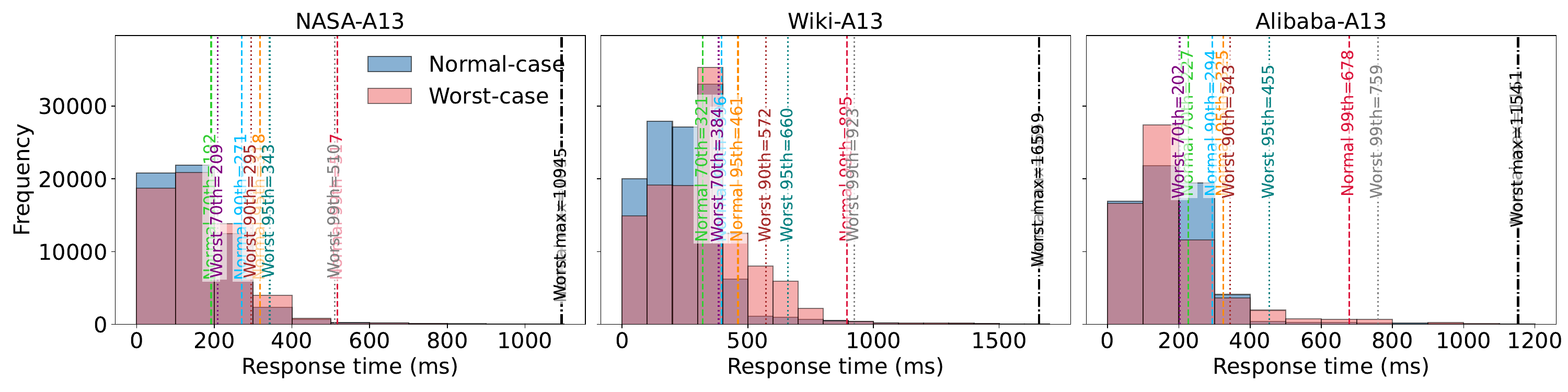}
    \caption{Response time distributions of normal-case and worst-case transient effects generated by HGraphScale.}
    \label{fig:worst_case}
\end{figure*}

Fig.~\ref{fig:worst_case} compares the \emph{normal-case scenario} and \emph{worst-case scenario} of A13 with respect to different traces. This figure demonstrates distributions and percentiles of response time. We can observe from this figure that the long tail response time of both \emph{normal-case scenario} and \emph{worst-case scenario} is similar in NASA-A13. As for Wiki-A13 and Alibaba-A13, although the 90th and 95th percentile response times in the worst-case scenario are about 200 ms higher than those in the normal-case, the differences in the 99th percentile response time and the maximum response time remain small (less than 100 ms). These results demonstrate that HGraphScale maintains strong performance and robustness even under worst-case transient effects. 

In summary, the worst-case scenario analysis shows that even when scaling operations are assumed to incur maximum transient overheads, HGraphScale consistently preserves stable performance. The response time distributions across different traces indicate that only moderate increases occur at the 90th and 95th percentiles, while the tail latency (99th percentile and maximum) remains largely unaffected. These findings confirm that HGraphScale delivers robust and resilient autoscaling decisions, ensuring bounded latency even under adverse conditions.

\section{Effectiveness of $\zeta_{con_c}$}
One of the container feature ($\zeta_{con_c}$) represents the remaining resources of the VM that hosts $con_c$. To further analyze the role of this feature, we conducted ablation studies by excluding it, producing a variant named \emph{w/o $\zeta_{con_c}$}. TABLE~\ref{table:ablation_feature} reports the ART results of \emph{w/o $\zeta_{con_c}$} and HGraphScale across five application structures under the Alibaba trace. We observe that \emph{w/o $\zeta_{con_c}$} performs slightly better than HGraphScale only on Alibaba-A13, while in all other scenarios it is significantly worse. These results highlight the importance of the remaining resources of VMs in making effective autoscaling decisions for containers.

\begin{table}[htb!]
    \centering
    \caption{Comparison results between w/o $\zeta_{con_c}$ and HGraphScale in terms of ART.}
    \label{table:ablation_feature}
    \begin{tabular}{c|c|c}
        \hline
        Scenario & w/o $\zeta_{con_c}$ & HGraphScale    \\
        \hline
        Alibaba-A11 & 237.70 & \textbf{222.17} \\
        \hline
        Alibaba-A12 & 292.53 & \textbf{283.78} \\
        \hline
        Alibaba-A13 & \textbf{175.45} & 178.91 \\
        \hline
        Alibaba-A14 & 362.19 & \textbf{299.28} \\
        \hline
        Alibaba-A30 & 437.02 & \textbf{183.94} \\
        \hline
    \end{tabular}
\end{table}

\section{Number of Replicas}
\begin{figure}[htb!]
    \centering
    \includegraphics[width=1\linewidth]{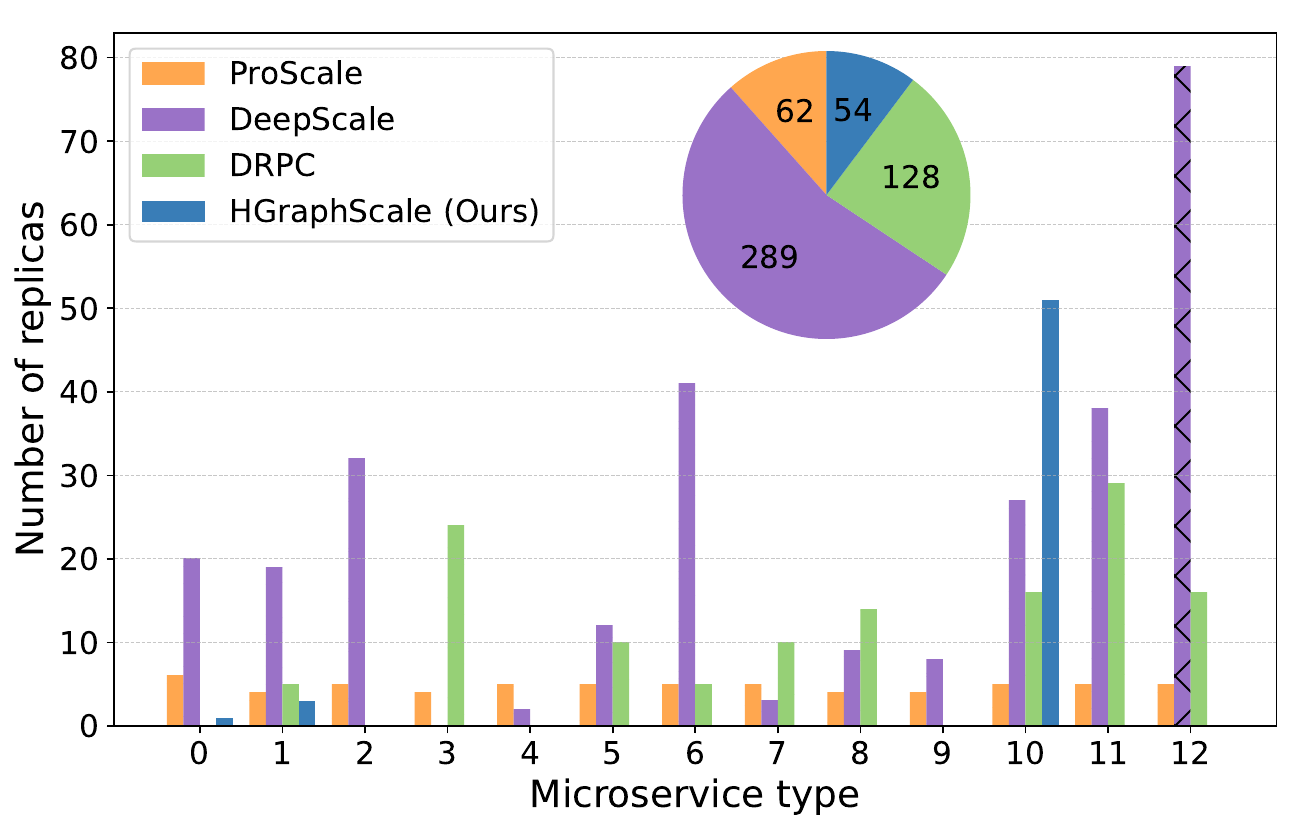}
    \caption{Number of replicas in NASA-13.}
    \label{fig:replica_number}
\end{figure}

Fig.~\ref{fig:replica_number} presents the replica numbers of each microservice type generated by different autoscaling approaches in NASA-13, as well as the total number of replicas produced by each approach, as shown in the pie chart. We can see that HGraphScale uses the least number of replicas when compared to DeepScale and DRPC, yet simultaneously achieving the fastest response time, as discussed in previous section.

\bibliographystyle{splncs04}
\bibliography{bibliography}